\documentclass[10pt]{ietbook}
\usepackage{graphicx}
\usepackage{adjustbox}
\usepackage{color}
\usepackage{amsmath}
\usepackage{algorithm}
\usepackage{pifont}
\usepackage{algpseudocode}
\usepackage[affil-it]{authblk}

\usepackage{etoolbox}
\patchcmd{\thebibliography}{\chapter*}{\section*}{}{}
\renewcommand\bibname{References}

\begin{document}
\setcounter{chapter}{13}
\markboth{SDN helps Big-Data to optimize access to data}{Y. Fu and F. Song}

%Book Title: Big Data and Software Defined Networks

\author{Yuankun Fu \footnote{Department of Computer Science, Purdue University, USA} and Fengguang Song \footnote{Department of Computer Science, Indiana University-Purdue University Indianapolis, USA}}
%\affil{Department of Computer Science, Indiana University-Purdue University Indianapolis}

\chapter{SDN helps Big-Data to optimize access to data}

This chapter introduces the state-of-the-art in the emerging area of combining High Performance Computing (HPC) with Big Data Analysis. To understand the new area, the chapter first surveys the existing approaches to integrating HPC with Big Data. %After the survey, the chapter will present a list of open questions. The list of open questions 
Next, the chapter introduces several optimization solutions that focus on how to minimize the data transfer time from computation-intensive applications to analysis-intensive applications as well as minimizing the end-to-end time-to-solution. The solutions utilize SDN to adaptively use both high speed interconnect network and high performance parallel file systems to optimize the application performance. 
A computational framework called DataBroker is designed and developed to enable 
a tight integration of HPC with data analysis. Multiple types of experiments have been conducted to show different performance issues in both message passing and parallel file systems and to verify the effectiveness of the proposed research approaches.

\section{Introduction}
%Introduction to the integration of High Performance Computing with Big Data Analytics
\label{sect:introduction}

Alongside experiments and theories, computational modeling/simulation and big data analytics
have established themselves as the critical {\it third} and {\it fourth} paradigms in modern scientific discovery
\cite{scientific-bd-2013, doe-synergistic}.
%Instead of researching into either computation-intensive applications or data-intensive applications
Nowadays, there is an inevitable trend towards integrating different applications of computation and data analysis together.
The benefits of combining them are significant:
(1) The overall end-to-end time-to-solution can be reduced
considerably such that interactive or real-time scientific discovery becomes feasible;
(2) the traditional one-way communication (from computation to analysis) can be bidirectional
to enable guided computational modeling and simulation;
and 3) computational modeling/simulation and data-intensive analysis are complementary to each other and can be used in a virtuous circle to amplify their collective effect.

%Challenges, followed by our proposed solutions!
%What are the difficulties? Research problem a few questions\
%predict time-to-solution for this type of problem, how to model the problem
%how to couple simulation and analysis efficiently
%how to handle all the bottlenecks, our solution will cover adaptive three situation
However, it is a challenging task to integrate computation with analysis effectively.
Critical questions include: How to minimize the cost to couple computation and analysis?
how to design an effective software system to enable and facilitate such an integration?
and how to optimize the co-scheduling of different computation and data-intensive applications?
In this chapter, we build an analytical model to estimate the overall execution time
of the integrated computation and data analysis, and design an intelligent data broker to
intertwine the computation stage and the analysis stage to achieve the optimal
time-to-solution predicted by the analytical model.

\begin{figure}[htbp]
    \begin{centering}
    \includegraphics[width=0.55\textwidth]{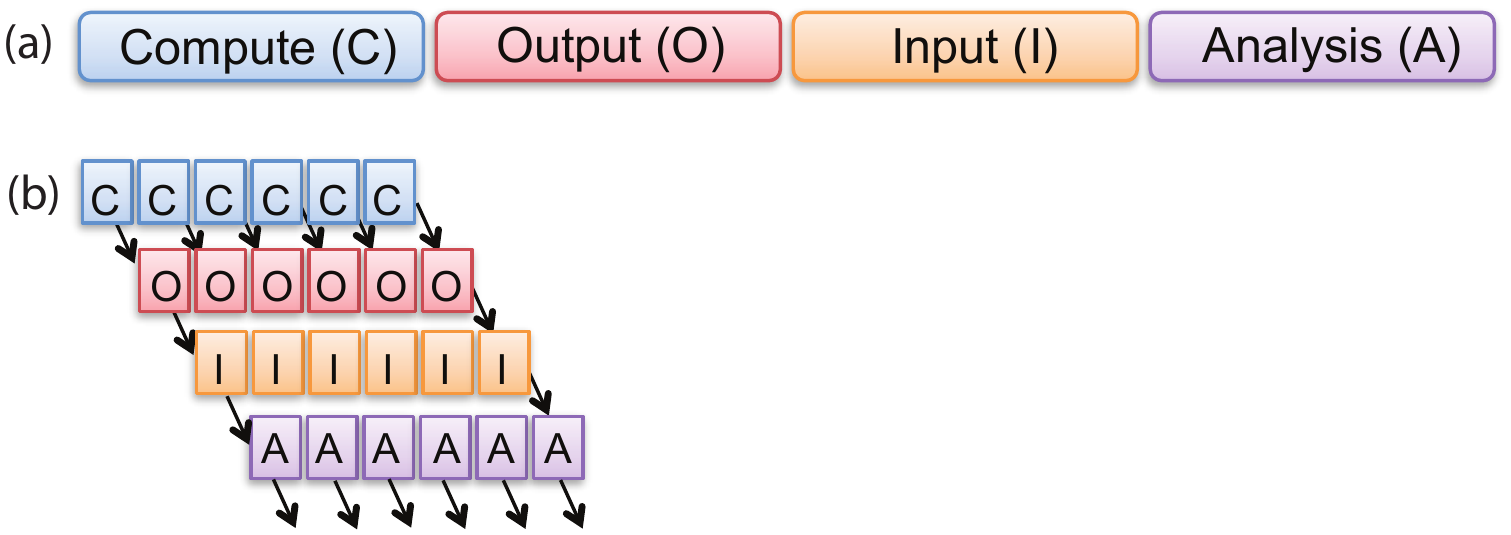}
    \caption{\small Comparison between the traditional process (a) and the new fully asynchronous pipeline method (b).}
    \label{fig:pipeline}
    \end{centering}
\end{figure}

To fully interleave computation with analysis, we propose and introduce a fine-grain-block task-based asynchronous
parallel execution model. The execution model utilizes the abstraction of pipelining, which is
widely used in computer architectures \cite{patterson2013computer}.
In a traditional scientific discovery,
a user often executes the computation, stores the computed results to disks,
then reads the computed results, and finally performs data analysis.
From the user's perspective, the total time-to-solution is the sum of the four execution times.
In this chapter, we rethink of the problem by using a novel method of fully asynchronous pipelining.
With the asynchronous pipeline method (detailed in Section \ref{sect:model}), a user input is divided into fine-grain blocks.
Each fine-grain block goes through four steps: computation, output, input, and analysis.
As shown in Figure \ref{fig:pipeline}, our new end-to-end time-to-solution is equal to the maximum of the
the computation time, the output time, the input time, and the analysis time (i.e., the time of a single step only) \cite{fu2016iccs}.
Furthermore, we build an analytical model to predict the overall time-to-solution
to integrate computation and analysis, which provides developers with
an insight into how to efficiently combine them.

Although the analytical model and its performance analysis reveal that
the corresponding integrated execution can result in high performance,
there is no software available to support the online tight coupling of analysis and computation at run time.
%and the new way of fine-grain asynchronous execution.
To facilitate the asynchronous integration of computation and analysis,
we design and develop an I/O middleware, named {\em Intelligent DataBroker},
to adaptively prefetch and manage data in both secondary storage and main memory to minimize the I/O cost.
This approach is able to support both in-situ (or in memory) data processing and
post-processing where initial dataset is preserved for the entire community (e.g., the weather community)
for subsequent analysis and verification.
The computation and analysis applications are coupled up through the DataBroker.
DataBroker consists of two parts: a DataBroker producer in the compute node to send data, and a DataBroker
consumer in the analysis node to receive data.
It has its own runtime system to provide dynamic scheduling, pipelining, hierarchical buffering, and prefetching. This chapter introduces the design of the current prototype of DataBroker briefly.

We performed experiments on BigRed II (a Cray system) with a Lustre parallel file system
at Indiana University to verify the analytical model and
compare the performance of the traditional process, an improved version of the traditional process
with overlapped data writing and computation, and our fully asynchronous pipeline approach.
Both synthetic applications and real-world computational fluid dynamics (CFD) applications %of computational fluid-structure interactions
have been implemented with the prototype of DataBroker.
Based on the experiments,
the difference between the actual time-to-solution and the predicted time-to-solution is less than 10\%.
Furthermore, by using DataBroker,
our fully asynchronous method is able to outperform the improved traditional method by up to 78\%
for the real-world CFD application.

In the remainder of the chapter,
Section \ref{sect:related-work} covers the state of the art.
Section \ref{sect:benchmark} compares the efficiency difference between the
message passing method and the parallel file I/O method for transferring data between applications.
Section \ref{sect:model} introduces the analytical model to estimate the time-to-solution.
Section \ref{sect:implementation} presents the DataBroker middleware to enable an optimized integration approach.
Section \ref{sect:results} verifies the analytical model and demonstrates the speedup using the integration approach.
Finally, 
Section \ref{sect:challenges} raises a few open questions, and
Section \ref{sect:conclusion} summarizes the chapter.
%and points out its future work.
 %budget: 1 page.

\section{State of the Art and Related Work}
\label{sect:related-work}
This section introduces the existing approaches to integrating computation with data analysis,
and compares our work with related work.

{\it The conventional approach}: 
As shown earlier in Figure \ref{fig:pipeline}, 
the computational modeling and simulation applications \cite{simulation-book-1, 2014unforeseen, madadgar2014towards} will compute and output computed results to files. 
Providing that the file format is known, any data analysis application can be launched to perform various types of data analysis. The advantage is that
independent software projects can be developed at the same time in separate organizations.
Also, the computed results can be stored and analyzed later by other analysis applications.
The problem is that it is strictly sequential, which results in long end-to-end time-to-solution.

{\it The in-situ approach}: %high performance, not good SE, not flexible, not realistic
In opposite directions of the conventional approach,
the in-situ approach analyzes data when the data are still resident in memory.
While the I/O cost is eliminated, it has three issues \cite{wong2012top, doe-synergistic, ma2009situ}:
1)  It takes a lot of effort to couple the computation code with the analysis code.
Developers have to place analysis functions to the address space of the modeling/simulation application,
which requires data-format conversion and good understanding of both computation and analysis domains;
2) Many real world applications are already tight on memory. %and are limited by the available memory,
Allocating memory to in-situ analysis (together with other resource contentions)
will slow down both computation and analysis processes. 
%Or they may have a severe load imbalance between their computational and analysis processes. 
Since analysis applications are typically less scalable than computation applications,
the computation applications will be stalled due to the limited in-situ memory space; 
and 3) 
It does not support preserving data for long-term studies and 
the entire community for different data analyses. 
%and the analysis must be known prior to the actual modeling and simulation.
%and verification. %for independently analyze and verify.

{\it The data-staging approach}: %Best, but not really coupled; everything is synchronous; 
Unlike the in-situ approach, which requires writing custom code and sharing the same resources,
a general approach is to use {\it data staging}
to place analysis on an analysis cluster or a separate partition of a supercomputer.
The data-staging approach has a flexible design, in which computation and analysis can execute
on the same compute node, on different compute nodes, or on different HPC systems.
The approach can also minimize resource contention (e.g., CPU, memory, bus, disk) between computation and analysis processes. 
To support transferring data from computation to analysis, a few I/O libraries and middleware have been implemented.
FlexIO \cite{flexio} uses both shared memory and RDMA to support data analysis 
either on the same or different compute nodes.
GLEAN \cite{glean} uses sockets to support data staging on analysis nodes on an analysis cluster.
%Other significant work such as DataStager \cite{datastager}, I/O Container \cite{iocontainer}, 
%and DataSpaces \cite{dataspaces} uses RDMA to 
DataStager \cite{datastager} and I/O Container \cite{iocontainer} use RDMA to
provide staging on a separate part of compute nodes. %to support data analysis.
DataSpaces \cite{dataspaces} provides a distributed virtual space and allows
different processes to put/get data tuples to/from the space.

%{\it Our approach and differences}: 
%paradigm-unifying, cross-layer, global scheduling, dataflow, fully async
We create a new computing framework called DataBroker that
is based on a data-driven data-staging service.
It provides a unifying approach,
which takes into account {\it computation}, {\it output}, {\it input}, and {\it analysis} as a whole,
and performs global dynamic scheduling across computation and analysis to optimize the end-to-end time-to-solution. 
This approach also builds analytical models to estimate the performance of the integrated computation and analysis and optimizes the time-to-solution. 

The new DataBroker is used to fasten analysis applications to computational applications.
Different from the existing staging middleware,
DataBroker can perform in-memory analysis or file-system based analysis---adaptively---without 
stalling the computation processes. 
Moreover,
DataBroker %allows different processes to write/read data blocks but 
focuses on fine-grain pipelining operations.   
There exist efficient parallel I/O libraries
such as MPI-IO \cite{mpi-io}, ADIOS \cite{adios, adios-hello}, Nessie \cite{nessie}, and PLFS \cite{plfs} to 
%enable 
allow applications to adapt their I/O to specific file systems. 
We do not compete with these work. DataBroker is in the application level, 
which can use these techniques to optimize DataBroker's performance.

 %budget: 1/4 page.

\section{Performance analysis of message passing and parallel file system I/O}
\label{sect:benchmark}
%Outline
%In this section, we will discuss the performance of two data transfer method. 
In a typical scientific workflow, there are two available data transfer methods.
One way is to transfer data via MPI library, 
which is a widely-used message-passing library on parallel computing systems.
The other is to use file system via writing and reading files. 
Nowadays, file system has evolved a lot and becomes much faster than before. 
Current novel HPC systems mostly utilize high speed SSD to store temporary data into local disk, meanwhile use parallel file system to store long-term data.

Naturally, one may think file system is slow to transfer data compared with MPI messages.
But with the rapid development of parallel file system and the emergence of systems equipped with SSD, 
it is necessary to reconsider this issue. 
Thus, we compare the performance of transferring data using MPI and file I/O.
In order to achieve the goal, we design two sets of experiments.
The first set is used to measure the time to transfer one block of data by one MPI message.
The second set is used to get the time to transfer one block of data using file I/O.

The first set of MPI experiments is designed as follows.
A producer process creates $n$ blocks filled with random values and
sends them to a consumer processes using $n$ MPI messages.
We measure the total time of MPI\_Send function on each thread
and assign it as the data transfer time $T_{MPI}$. 
Thus, the time to transfer a data block by MPI is $T_{MPI/Block} = \frac{T_{MPI}}{n}$. 
%Then we vary the block size to get $T_{MPI/Block}$ on different block sizes.
% Each configuration experiment will be conducted several times and we get the average.

The second set of file I/O experiments is described as follows.
A producer process creates $n$ blocks filled with random values and
writes them to disk. 
After the producer process has finished, a consumer process will start reading the files. %one by one 
We measure the total writing time $T_{Write}$ and reading time $T_{Read}$ on each thread
and use their sum as the data transfer time.
Thus, the time to transfer a data block by parallel file I/O is 
$T_{HDD/Block}=\frac{T_{Write}+T_{Read}}{n}$.
%Accordingly, the Comet SSD transfer is $T_{SSD/Block} = \frac{T_{SSD\_Write}+T_{SSD\_Read}}{n}$.
%Then we still vary the block size to get $T_{HDD/Block}, T_{SSD/Block}$ on different block sizes.

The two sets of experiments are performed on BigRed II and Comet. 
BigRed II is a Cray XE6/XK7 HPC system in Indiana University. 
%The compute nodes are connected by a Gemini interconnect network. 
It contains a Gemini interconnect network and a Lustre parallel file system named Data Capacitor II (DC2). %shared by all users. 
Comet is a dedicated XSEDE cluster in San Diego Supercomputer Center.
It contains hybrid fat-tree interconnect network and a Lustre parallel file system.
Besides, each Comet compute node has 320 GB SSD local storage. 
Thus, we conduct the two sets of experiments first on BigRed II to get $T_{MPI/Block}$ and $T_{HDD/Block}$,
and then on Comet to get $T_{MPI/Block}$, $T_{HDD/Block}$ and $T_{SSD/Block}$.

\begin{figure}[htbp]
	\centering
	\begin{minipage}[t]{2.3in}
		\centering
		\includegraphics[width=1\textwidth]{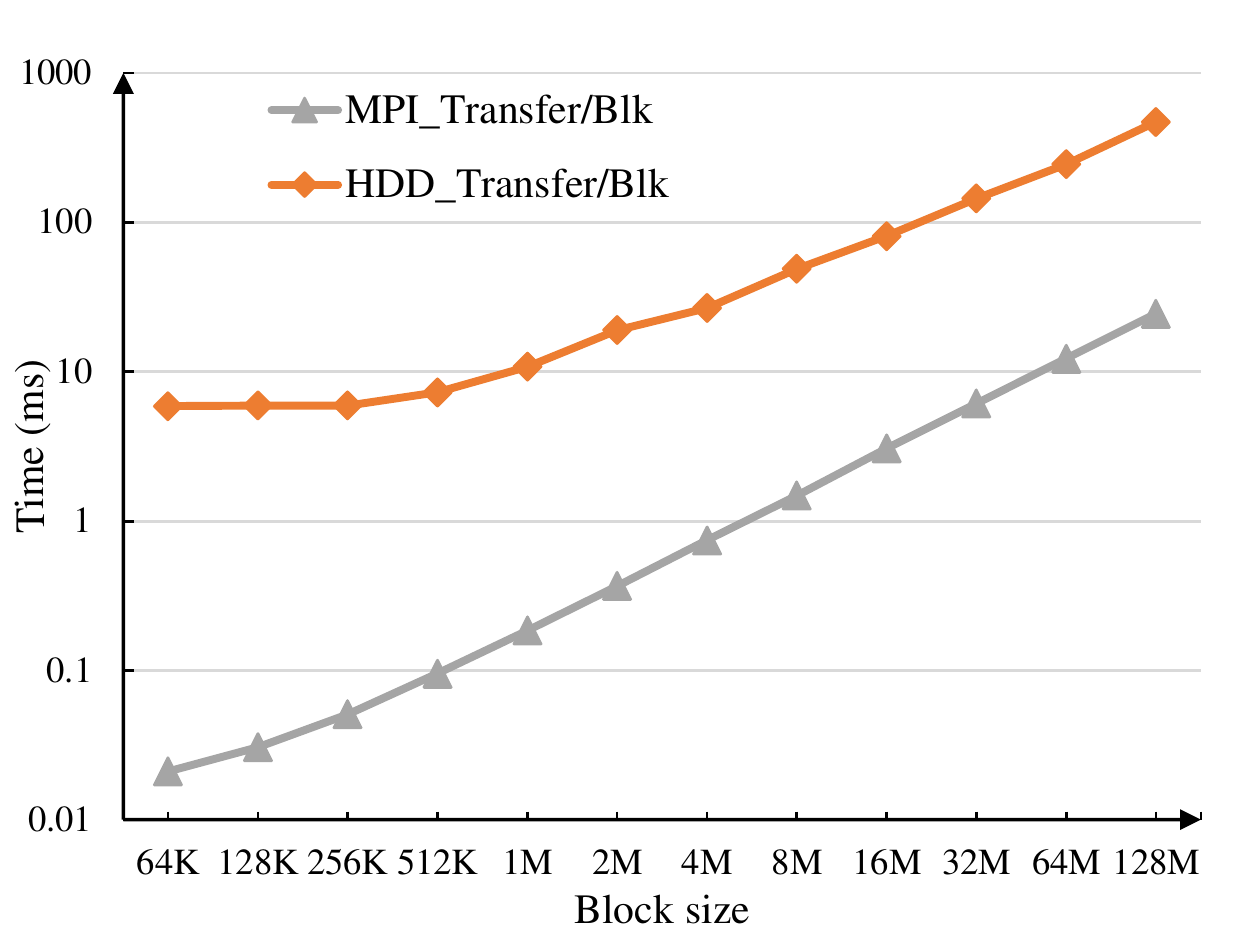}
		\par \centering {\small (a) Data Transfer Time via MPI and HDD}
		\label{fig:sub1}
	\end{minipage}
	\hfill
	\begin{minipage}[t]{2.3in}
		\centering
		\includegraphics[width=1\linewidth]{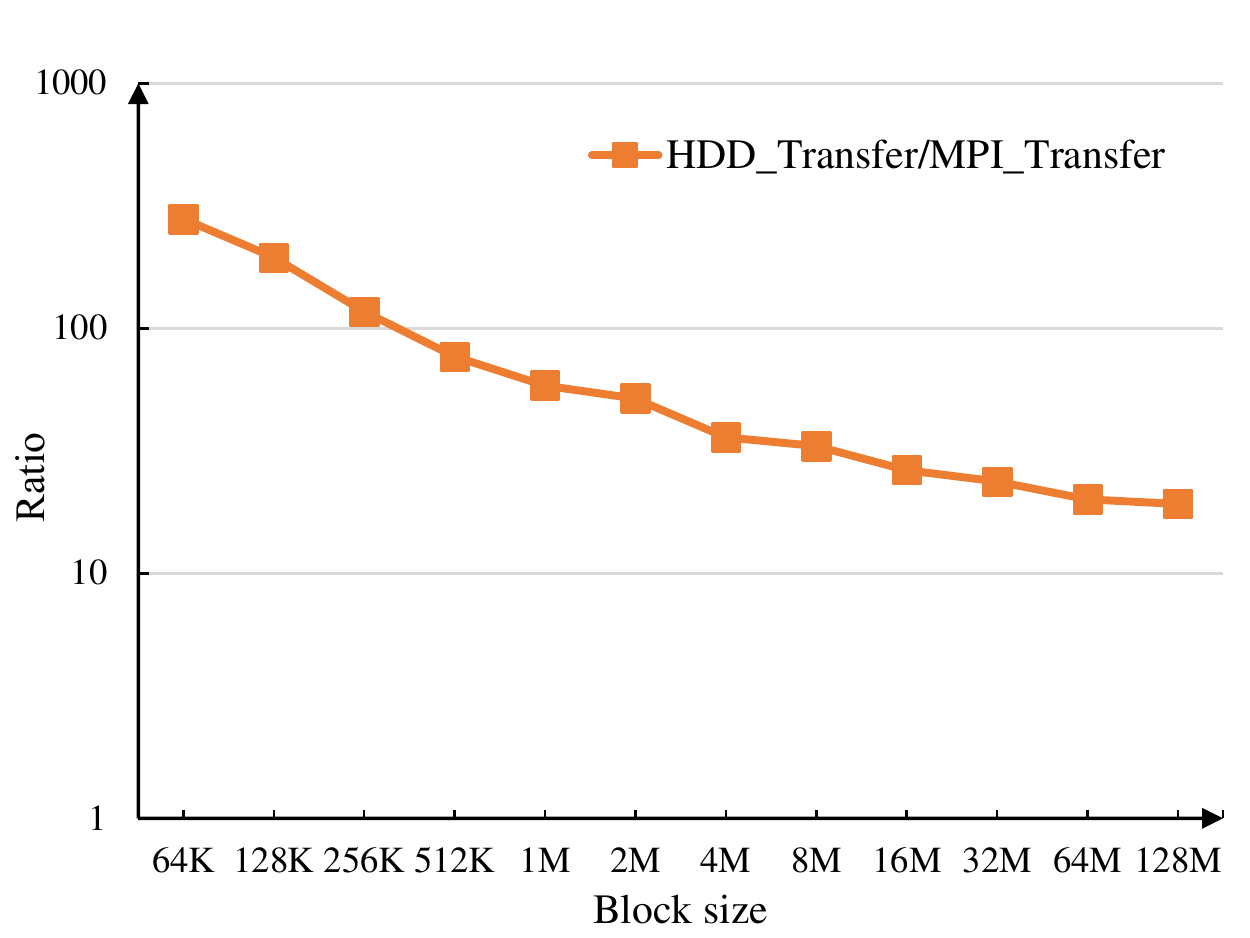}
		\par \centering {\small (b) Ratio of MPI/HDD}
		\label{fig:sub2}
	\end{minipage}
	\caption{\small Data transfer Time and speedup for one block data on BigRed II.}
	\label{fig:BG2-performance-MPI-HDD}
\end{figure}

Figure \ref{fig:BG2-performance-MPI-HDD} (a) shows the result of $T_{MPI/Block}$ and $T_{HDD/Block}$ on BigRed II. 
We can observe that MPI is faster than parallel file I/O on all block sizes.
In addition, both $T_{MPI/Block}$ and $T_{HDD/Block}$ will increase as the block size increases. 
However, the performance gap between MPI and parallel file I/O becomes gradually narrower.
We use the ratio of $\frac{T_{HDD/Block}}{T_{MPI/Block}}$ to measure it. 
Figure \ref{fig:BG2-performance-MPI-HDD} (b) shows that MPI outperforms parallel file I/O by 278 times on 64KB,
but the ratio falls to 19 times on 128MB. 
This result reflects that MPI is excellent in transferring data on small block size,
but the benefit loses as block size grows larger. 
%So there is chance to use disk when data block size is larger than 1MB.

\begin{figure}[htbp]
	\centering
	\begin{minipage}[t]{2.3in}
		\centering
		\includegraphics[width=1\textwidth]{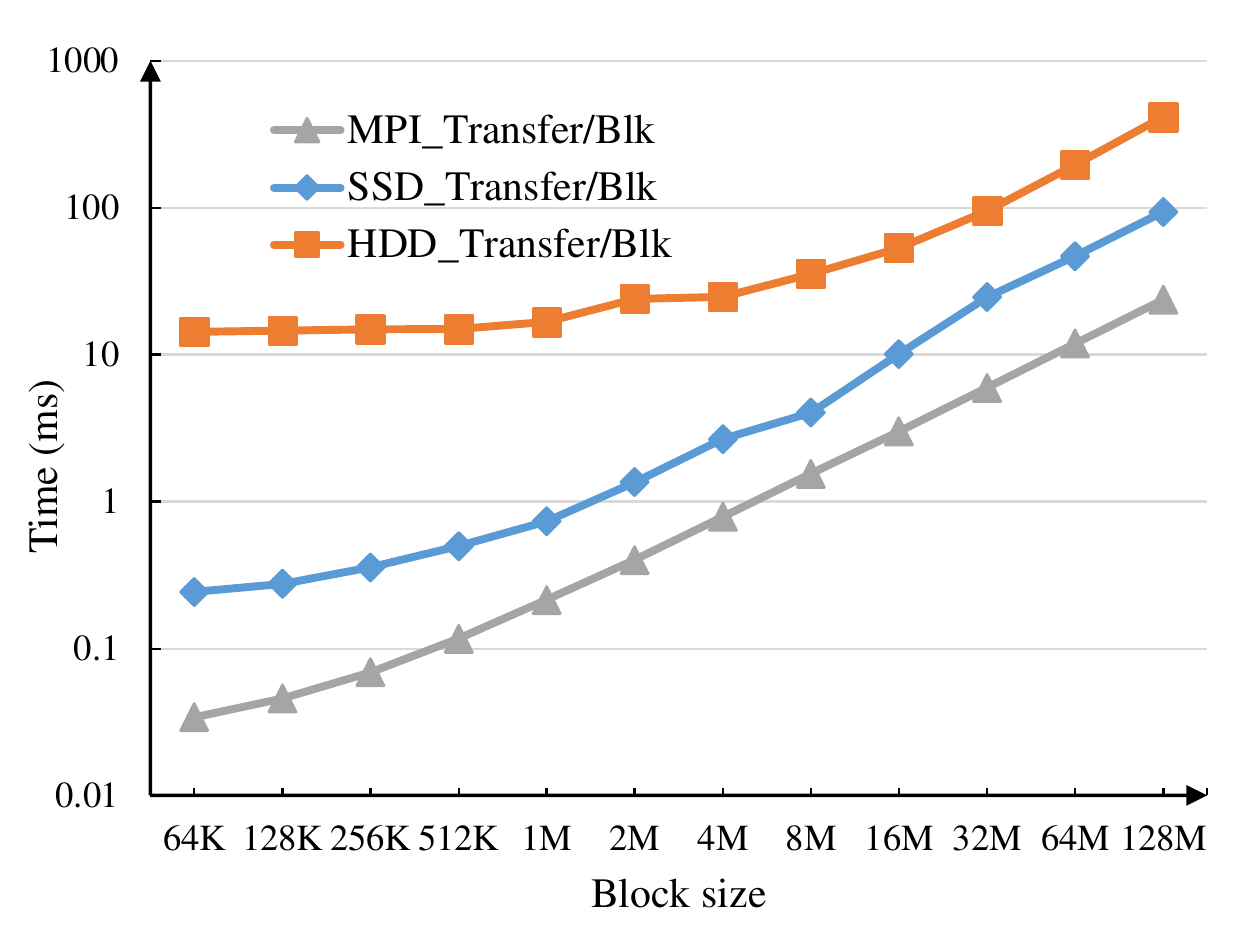}
		\par \centering {\small (a) Data Transfer Time via MPI, SSD and HDD}
		\label{fig:sub1}
	\end{minipage}
	\hfill
	\begin{minipage}[t]{2.3in}
		\centering
		\includegraphics[width=1\linewidth]{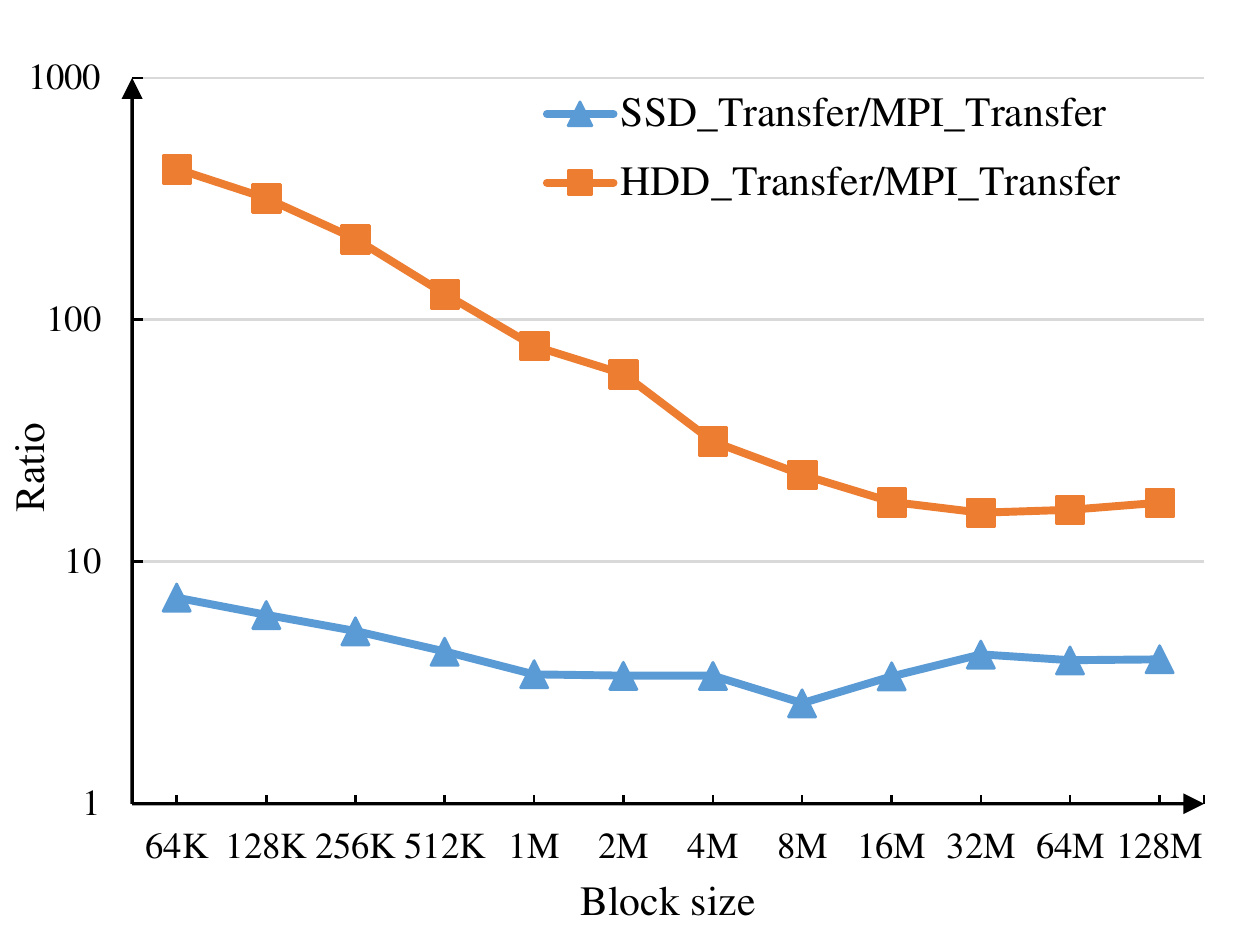}
		\par \centering {\small (b) Ratio of MPI/SSD and MPI/HDD}
		\label{fig:sub2}
	\end{minipage}
	\caption{\small Data transfer Time and speedup for one block data on Comet.}
	\label{fig:comet-performance-MPI-SSD-HDD}
\end{figure}

Figure \ref{fig:comet-performance-MPI-SSD-HDD} (a) shows the results on Comet.
%$T_{MPI/Block}$, $T_{HDD/Block}$ and $T_{SSD/Block}$  
We can find that the time to transfer a data block using MPI is faster than using local SSD, 
and using parallel file system is the slowest.
Moreover, $T_{MPI/Block}$, $T_{SSD/Block}$ and $T_{HDD/Block}$ increase as block size increases.
Again, we find that the performance gap among them becomes narrower.
We still use the ratio of $\frac{T_{HDD/Block}}{T_{MPI/Block}}$ and $\frac{T_{SSD/Block}}{T_{MPI/Block}}$ to measure the trend.
Figure \ref{fig:comet-performance-MPI-SSD-HDD} (b) shows that 
MPI outperforms parallel file I/O by 419 times on 64KB,
but the ratio drops to 17 times on 128MB.
On the other hand, MPI outperforms SSD by 2.6 times on 8MB, and up to 7 times on 64KB.
%$\frac{T_{SSD/Block}}{T_{MPI/Block}}$ ranges from 2.6 on 8MB to 7 on 64KB.
To transfer 128MB data blocks, MPI is 2.9 times faster than SSD. %3.9-1=2.9
This suggests that using SSD to transfer both small and large blocks is an acceptable choice.

%Can use lfs parameter -stripesize -count OST to tune performance.

%Discussion:
%At last, we want to give a conclusion that MPI message passing is good for transfer small block size and slow to transfer big file size, while parallel file system is not that bad when transferring big file size. Thus file system is useful sometimes.

From the above experiment results, we can summarize that
MPI message is good at transferring small data size,
but its performance drops when transferring larger data size.
Moreover, on both machines, we find that MPI renders at most 19 times better performance when block size increases to 128MB. 
Thus, parallel file system is not that slow in the case of transferring large file. Especially when equipped with SSD,
file system will be helpful in transferring both small and large data size. 
%Our future work includes combining MPI and file I/O to create a dual-path data transfer method.

 %budget: 1.5 page.

\section{Analytical modelling based end-to-end time optimization}
\label{sect:model}
\subsection{The Problem}
This chapter targets an important class of scientific discovery applications which require
combining extreme-scale computational modeling/simulation with large-scale data analysis.
The scientific discovery consists of computation, result output, result input, and data analysis.
From a user's perspective, the actual time-to-solution is the end-to-end time from the start of the computation
to the end of the analysis.
While it seems to be a simple problem with only four steps, different methods to execute the four
steps can lead to totally different execution time.
For instance, traditional methods execute the four steps sequentially such that the overall time-to-solution is the sum of the four times.

In this section, we study how to unify the four seemingly separated steps into a single problem and build
an analytical model to analyze and predict how to obtain optimized time-to-solution.
The rest of the section models the time-to-solution for three different methods:
1) the traditional method, 2) an improved version of the traditional method,
and 3) the fully asynchronous pipeline method.

\subsection{The Traditional Method}
\begin{figure}[htbp]
\centering
\includegraphics[width=0.52\textwidth]{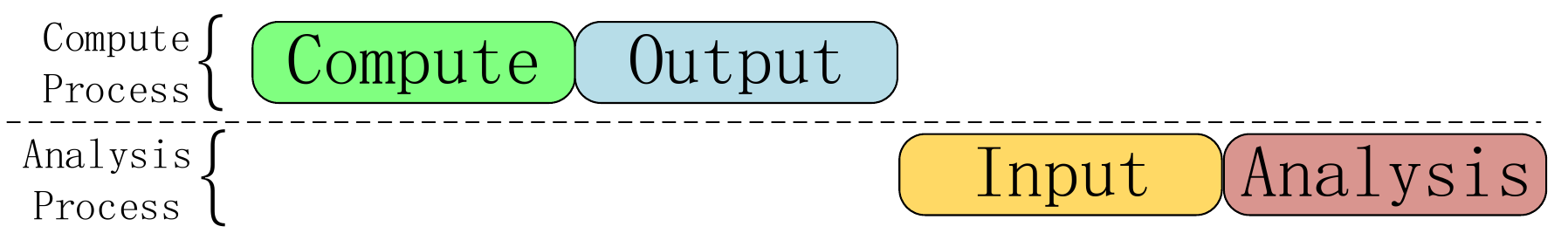}
\caption{\small The traditional method.}
\label{fig:exp0}
\end{figure}

Figure \ref{fig:exp0} illustrates the traditional method, which is the simplest method without optimizations
(next subsection will show an optimized version of the traditional method) \cite{fu2016iccs}.
The traditional method works as follows: the compute processes compute results and write computed results to disks,
followed by the analysis processes reading results and then analyzing the results.

The time-to-solution ($t2s$) of the traditional method can be expressed as follows:
\begin{align*}
T_{t2s} = T_{comp} + T_{o} + T_{i} + T_{analy}
\end{align*}
,where $T_{comp}$ denotes the parallel computation time,
$T_{o}$ denotes the output time, $T_{i}$ denotes the input time,
and $T_{analy}$ denotes the parallel data analysis time.
%To quantify traditional model 0, we use some notations which will also be quoted later: $T_{comp}$ is the total time for compute nodes to compute the total scientific data; $T_{o}$ is the total time for compute nodes to write data to disk; $T_{i}$ is the total time for analysis nodes to read data from disk; $T_{analy}$ is the total time for analysis nodes to analyse data. $T_{sim}$ is total time of each compute node, while $T_{anl}$ is total time of each analysis node.$T_{t-to-s}$ means the total time to solution, which is the sum of the four parts.
%We define the total data size is $D$. We use $P$ cores to compute and $Q$ cores to analysis. $B$ is block size. Thus, the total blocks to be produced is $N=D/B$. Then each compute core will get $N/P=D/(B \cdot P)$ blocks while each analysis node will get $N/Q=D/(B \cdot Q)$ blocks.$t_{comp}$, $t_{o}$, $t_{i}$, $t_{analy}$ is time for each core to compute, output, input and analyse one block. Because each core to compute and analysis can run in parallel, $T_{comp}$ equals to each compute core to finish its job. The same goes with $T_{o}$, $T_{i}$ and $T_{ana}$. Then we get the formula below.\\
%\[T_{sim}=T_{comp}+T_{o}, T_{comp}= t_{comp} \cdot \frac{D}{B \cdot P},T_{o}= t_{o} \cdot \frac{D}{B \cdot P}\]
%\[T_{anl}=T_{i}+T_{analy}, T_{i}= t_{i} \cdot \frac{D}{B \cdot Q}, T_{analy}= t_{analy} \cdot \frac{D}{B \cdot Q}\]
%\[T_{t-to-s}=T_{sim}+T_{anl}=T_{comp}+T_{o}+T_{i}+T_{analy}\]
Although the traditional method can simplify the software development work,
this formula reveals that the traditional model can be as slow as the accumulated time of all the four stages.
%has two main disadvantages: firstly, it uses the whole data to operate rather than slice it into small blocks, which would cause cache and page replacement cost; secondly, it does not have any overlap between the four parts,which will get the sequential sum.

\subsection{Improved Version of the Traditional Method}
\begin{figure}[htbp]
\centering
\includegraphics[width=0.75\textwidth]{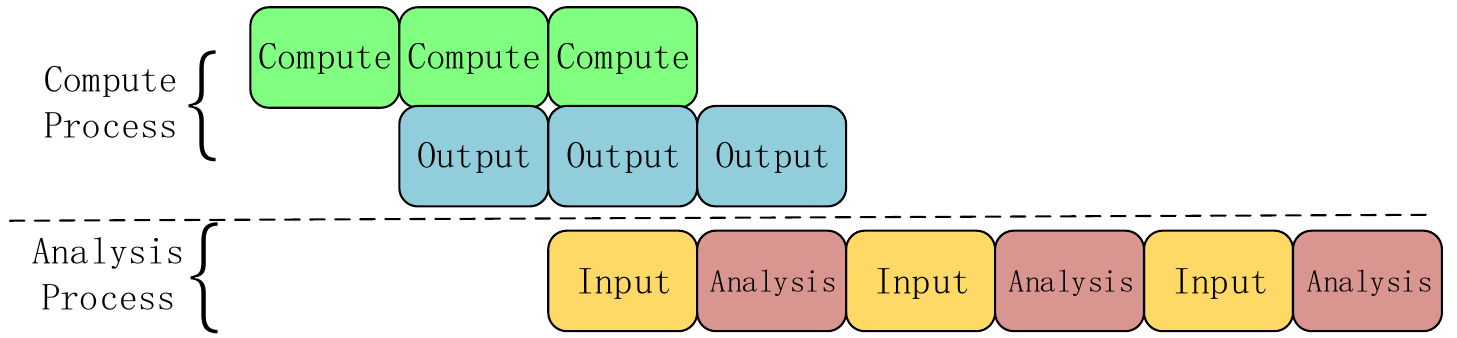}
\caption{\small An improved version of the traditional method.}
\label{fig:exp1}
\end{figure}

The traditional method is a strictly sequential workflow.
However, it can be improved by using multi-threaded I/O libraries, where
I/O threads are deployed to write results to disks meanwhile new results are generated by the compute processes.
The other improvement is that the user input is divided into a number of fine-grain blocks
and written to disks asynchronously.
Figure \ref{fig:exp1} shows this improved version of the traditional method \cite{fu2016iccs}.
We can see that the output stage is now overlapped with the computation stage
so that the output time might be hidden by the computation time.

%Tradition model 1 solves parts of disadvantages of traditional model 0. It uses block algorithm to slice the whole data into smaller blocks. Then it uses pipeline to integrate compute and output within compute nodes. But it still needs to sequentially input and analysis within analysis nodes. Figure \ref{fig:exp1} shows the traditional model 1.

Suppose a number of $P$ CPU cores are used to compute simulations and a number of $Q$ CPU cores
are used to analyze results, and the total amount of data generated is $D$.
Given a fine-grain block of size $B$, there are $n_b=\frac{D}{B}$ blocks.
Since scalable applications most often have good load balancing, we assume that
each compute core computes $\frac{n_b}{P}$ blocks and each analysis core analyzes $\frac{n_b}{Q}$ blocks.
The rationale behind the assumption of load balancing is that a large number of fine-grain parallel tasks
(e.g., $n_b \gg P$)
will most likely lead to an even workload distribution among a relatively small number of cores.

Our approach uses the time to compute and analyze individual blocks to estimate
the time-to-solution of the improved traditional method.
Let $t_{comp}$, $t_o$, $t_i$, and $t_{anal}$ denote the time to compute a block, write a block, read a block, and analyze a block, respectively.
Then we can get the parallel computation time $T_{comp} = t_{comp} \times \frac{n_b}{P}$,
the data output time $T_{o} = t_{o} \times \frac{n_b}{P}$,
the data input time $T_{i} = t_{i} \times \frac{n_b}{Q}$, and
the parallel analysis time $T_{analy} = t_{analy} \times \frac{n_b}{Q}$.
The time-to-solution of the improved version is defined as follows:
\begin{align*}
T_{t2s} = \max(T_{comp},T_{o},T_{i}+T_{analy})
\end{align*}
The term  $T_{i}+T_{analy}$ is needed because the analysis process
still reads data and then analyzes data in a sequence.
Note that this sequential analysis step can be further parallelized,
which results in a fully asynchronous pipeline execution model (see the following subsection).

%To quantify traditional model 1, we use the notation above. Because there is pipeline in simulation nodes, then $T_{sim}$ depends on the longer time between $T_{comp}$ and $T_{o}$. For example, if $t_{comp} \ge t_{o}$, because of the pipeline feature, each $t_{o}$ will hide under each $t_{comp}$, thus $T_{sim}$ will be the sum of $N/P\cdot t_{comp}$ plus the time $t_{o}$ for the final block to output. Then we get the formula below.

%\[\begin{split}
%T_{sim} = \left\{
%  \begin{array}{lr}
%    T_{comp}+t_{o}=t_{comp}\cdot \frac{D}{B \cdot P}+t_{o} & , t_{comp} \ge t_{o}\\
%    T_{o}+t_{comp}=t_{o}\cdot \frac{D}{B \cdot P}+t_{comp} & , t_{comp} < t_{o}\\
%  \end{array}
%\right.
%\end{split}\]

%Because analysis nodes does not have pipeline, we get the total time for analysis nodes as follows.
%\[T_{anl} = T_{i}+T_{analy}=(t_{i}+t_{analy})\cdot \frac{D}{B \cdot Q} \]
%Because there is also a pipeline between compute nodes and analysis nodes, when the total number of blocks $N$ is big enough, the total time-to-solution can be as follows:
%\[T_{t-to-s} \approx Max(T_{comp},T_{o},T_{i}+T_{analy})\]
%Because this is a pipeline model, we can interpret from the figure 2 that the time-to-solution equals to the longest part in computation, output and the sum of input and analysis.

\subsection{The Fully Asynchronous Pipeline Method}
\begin{figure}[tb]
\centering
\includegraphics[width=0.6\textwidth]{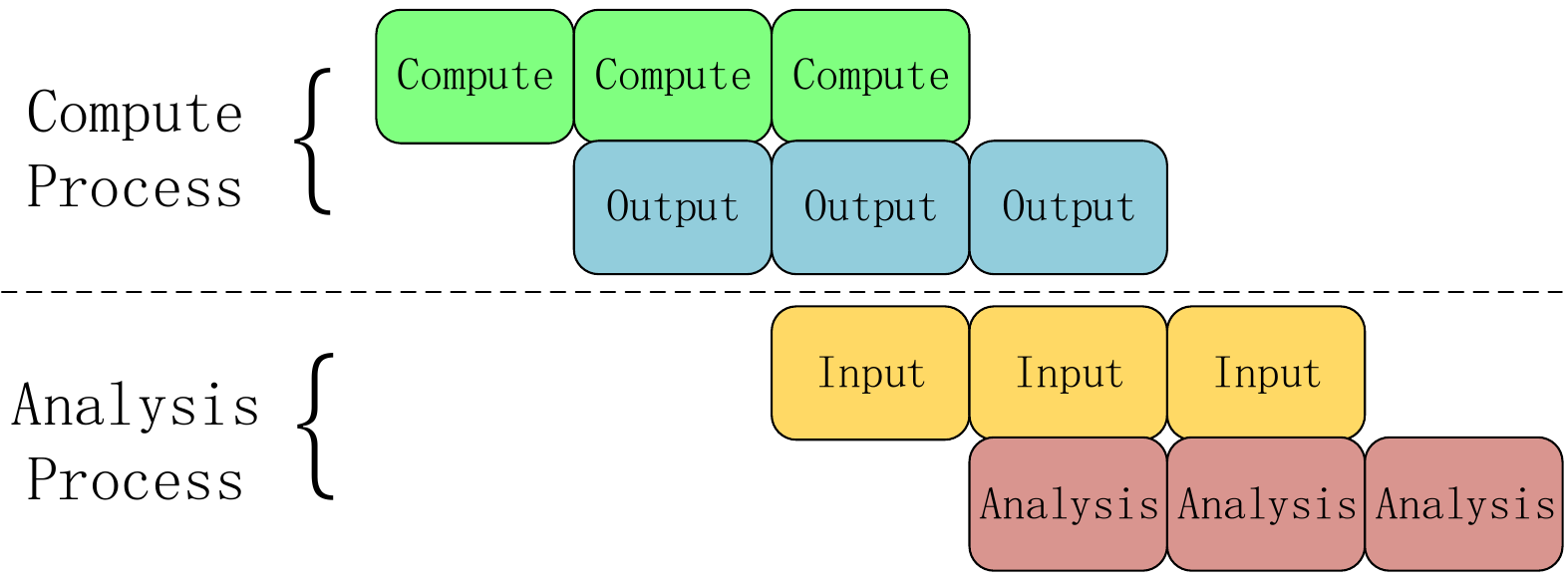}
\caption{\small The fully asynchronous pipeline method.}
\label{fig:exp2}
\end{figure}

The fully asynchronous pipeline method is designed to completely
overlap computation, output, input, and analysis such that
the time-to-solution is merely one component, which
is either computation, data output, data input, or analysis.
Note that the other three components will not be observable in the end-to-end time-to-solution.
As shown in Figure \ref{fig:exp2}, every data block goes through
four steps: compute, output, input, and analysis \cite{fu2016iccs}.
Its corresponding time-to-solution can be expressed as follows:
%\begin{ceqn}
\begin{align*}
%\begin{eqnarray*}
T_{t2s} &= \max(T_{comp}, T_{o}, T_{i}, T_{analy})\\
        &= \max(t_{comp}\times \frac{n_b}{P}, t_{o}\times \frac{n_b}{P}, t_{i}\times \frac{n_b}{Q}, t_{analy}\times \frac{n_b}{Q})
%\end{eqnarray*}
\end{align*}
%\end{ceqn}
The above analytical model provides an insight into how to achieve an optimal time-to-solution.
When $t_{comp}=t_{o}=t_{i}=t_{analy}$, the pipeline is able to proceed without any stalls and deliver
the best performance possible.
On the other hand, the model can be used to allocate and schedule computing resources to different stages appropriately to attain the optimal performance.

%This explains the pipeline model, the total time-to-solution equals to the longest time in the four parts.When $t_{comp}=t_{o}=t_{i}=t_{analy}$, we can get the best performance, that is, 1/4 of original time.\\
%Traditional model 1 has already couple compute, output and input into a pipeline through block algorithm. But we can find that input and analysis are still not coupled into the whole pipeline. Then we try to use prefetching technique to achieve a fully-asynchronous pipeline for input and analysis to improve the total performance. Figure \ref{fig:exp2} presents the idea.

%For compute nodes, the expected total time of computed nodes is the same as traditional model 1.
%\[\begin{split}
%T_{sim} = \left\{
%  \begin{array}{lr}
%    T_{comp}+t_{o}=t_{comp}\cdot \frac{D}{B \cdot P}+t_{o} & , t_{comp} \ge t_{o}\\
%    T_{o}+t_{comp}=t_{o}\cdot \frac{D}{B \cdot P}+t_{comp} & , t_{comp} < t_{o}\\
%  \end{array}
%\right.
%\end{split}\]
%When N is big enough, $T_{sim} \approx Max(T_{comp},T_{o})$.
%For analysis nodes, the expected total time of analysis nodes is as follows.
%\[\begin{split}
%T_{anl} = \left\{
%  \begin{array}{lr}
%    T_{i}+t_{analy}=t_{i}\cdot \frac{D}{B \cdot Q}+t_{analy} & , t_{i} \ge t_{analy}\\
%    T_{analy}+t_{i}=t_{analy}\cdot \frac{D}{B \cdot Q}+t_{i} & , t_{i} < t_{analy}\\
%  \end{array}
%\right.
%\end{split}\]
%When N is big enough, $T_{anl} \approx Max(T_{i},T_{analy})$.
%The total time-to-solution depends on which nodes complete at last.
%Thus, when N is big enough, we can get the following formula
%\[T_{t-to-s} \approx Max(T_{sim},T_{anl})=Max(T_{comp},T_{o},T_{i},T_{analy})\]
%\[T_{t-to-s} \approx Max(t_{comp}\cdot \frac{D}{B \cdot P},t_{o}\cdot \frac{D}{B \cdot P},t_{i}\cdot \frac{D}{B \cdot Q},t_{analy}\cdot \frac{D}{B \cdot Q})\]
%This explains the pipeline model, the total time-to-solution equals to the longest time in the four parts.When $t_{comp}=t_{o}=t_{i}=t_{analy}$, we can get the best performance, that is, 1/4 of original time.\\

\subsection{Microbenchmark for the Analytical Model}
%Ountline
To apply the analytical model in practice to predict the end-to-end time-to-solution, 
we need to know $t_{comp}, t_{o}, t_{i}$ and $t_{analy}$ on one block respectively. 
Given an application's block size, 
users can get $t_{comp}$ and $t_{analy}$ simply by running their sequential kernel of computation and analysis on one block.
The next step is to get $t_{o}$ and $t_{i}$.

At first, we design a naive microbenchmark to estimate $t_{o}$ and $t_{i}$.
The microbenchmark contains $P$ writer threads and $Q$ reader threads.
The mapping between writer threads and reader threads is static.
Given a block size, 
each writer thread creates $n$ blocks filled with random values and
writes them to disk.
After all the $P$ writer threads have completed,
the $Q$ reader threads start reading the files one by one. 
We measure the total time of writing files as $T_o$ and the total time of reading files as $T_i$ on each thread.
Thus, we get $t_{o}=\frac{T_o}{n}$ and $t_{i}=\frac{T_i \times Q}{P \times n}$.
%Hence, $T_{o}$ and $T_{i}$ are not easy variable to get. 

\begin{figure}[htbp]
	\centering
	\includegraphics[width=0.75\textwidth]{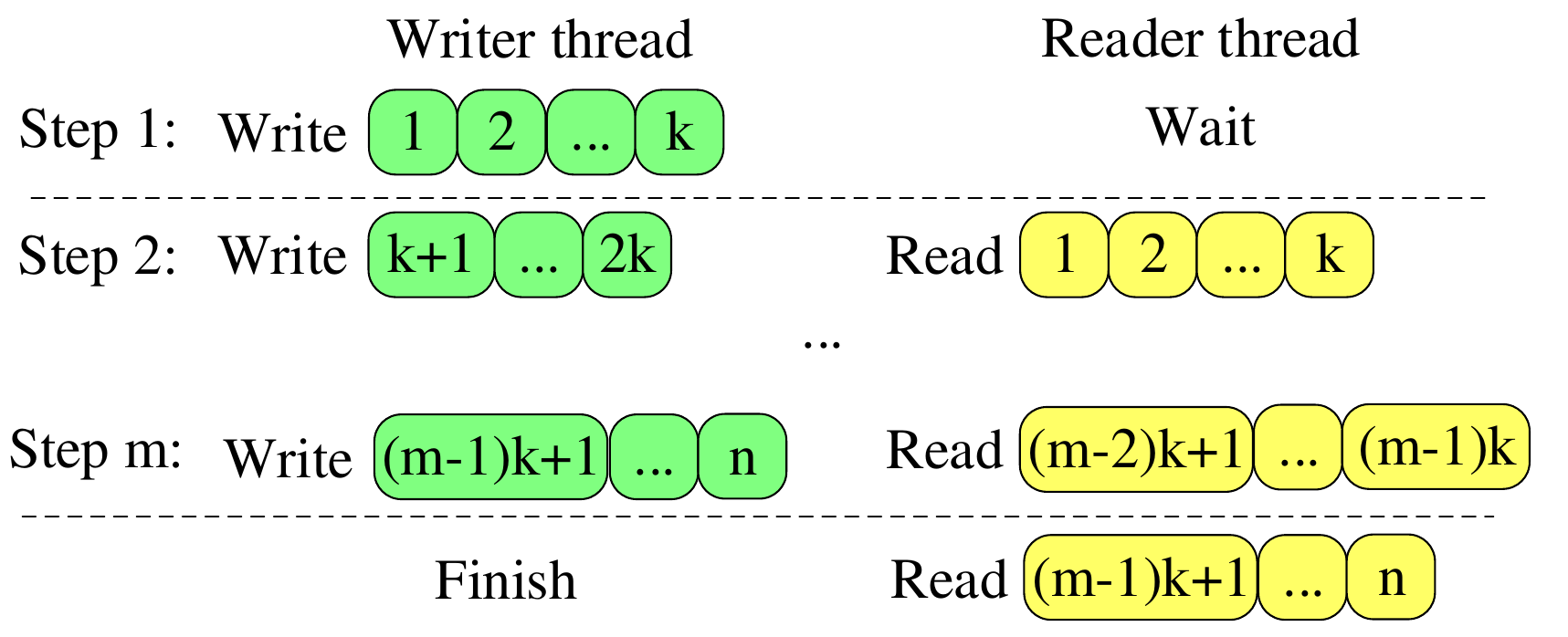}
	\caption{\small The Barrier Microbenchmark.}
	\label{fig:mb_alg}
\end{figure}

% naive not good, so we design microbenchmark
However, the above naive microbenchmark does not consider the scenario where multiple writes and reads can execute concurrently and asynchronously.
Thus, we propose a new version microbenchmark.
Figure \ref{fig:mb_alg} illustrates its idea,
each writer thread will generate $n$ blocks in $m$ steps.
Thus $k=\frac{n}{m}$ blocks are written into disk in each steps. 
Reader threads wait in the first step and start reading files in the second step,
and then reads the blocks generated by the writers in the previous step.

%For example, 
%in step 1, writer threads write blocks in the first step, and reader thread is idle.
%In step 2, writer threads write blocks in the second step, and reader threads read blocks in the first step.
%In step i, writer threads write blocks in the $i$th step, and reader threads read blocks in the $i-1$th step.
%We control the synchronization between steps using MPI\_Barrier().
To achieve this idea, we use MPI\_Barrier to synchronize between writer and reader threads.
As shown in Algorithm \ref{alg:writer} and \ref{alg:reader}, 
at the end of each step, all writers and readers will perform MPI\_Barrier to control $k=n/m$ blocks to be written and read. 
With this approach, we can simulate the scenario that file writing and reading executed at the same time.
\begin{algorithm}[htbp]
	\caption{Writer thread in new microbenchmark}\label{euclid}
	\begin{algorithmic}[1]
%		\Procedure{writer thread}{}
%		\State int $blk\_id\_begin$, $blk\_id\_ end$;
		\For {$i = 0$ to $m$}
			\State $blk\_id\_begin=i*n/m$
			\State $blk\_id\_end=(i+1)*n/m$
			\For {$blk\_id = blk\_id\_begin$ to $blk\_id\_end$}
				\State Write the block of $blk\_id$ to disk
			\EndFor
			\State MPI\_Barrier
		\EndFor
%		\EndProcedure
	\end{algorithmic}
	\label{alg:writer}
\end{algorithm}

\begin{algorithm}[htbp]
	\caption{Reader thread in new microbenchmark}\label{euclid}
	\begin{algorithmic}[1]		
%		\Procedure{reader thread}{}
		\State MPI\_Barrier
		\For {$i = 1$ to $m$}
		\State $blk\_id\_begin=(i-1)*n/m$
		\State $blk\_id\_end=i*n/m$
		\For {$blk\_id = blk\_id\_begin$ to $blk\_id\_end$}
		\State Read the block of $blk\_id$ from the mapped writer processes
		\EndFor
		\If{$i=m$} break
		\EndIf
		\State MPI\_Barrier
		\EndFor
%		\EndProcedure
	\end{algorithmic}
	\label{alg:reader}
\end{algorithm}

%Experiment to give $T_{o}$ and $T_{i}$ by different configurations. (1v1, 4v1, 16v4, some accurate experiments)
Then we perform experiments by different $P$ and $Q$ configurations on BigRed II. 
After we get the $t_{o}$ and $t_{i}$ from the two microbenchmarks, 
we compare them with the $t_{o}$ and $t_{i}$ obtained from the real application.
% combined with our DataBroker framework which is introduced in Section \ref{sect:implementation}.
At last, we use the relative error of writing and reading 
(i.e.
$\frac{|{t_{Real\_app}-t_{Naive\_MB}}|}{t_{Real\_app}}$ 
and $\frac{|{t_{Real\_app}-t_{Barrier\_MB}}|}{t_{Real\_app}}$)
to reflect the accuracy for each microbenchmark.

\begin{figure}[h]
	\centering
	\begin{minipage}[t]{2.3in}
		\centering
		\includegraphics[width=1\textwidth]{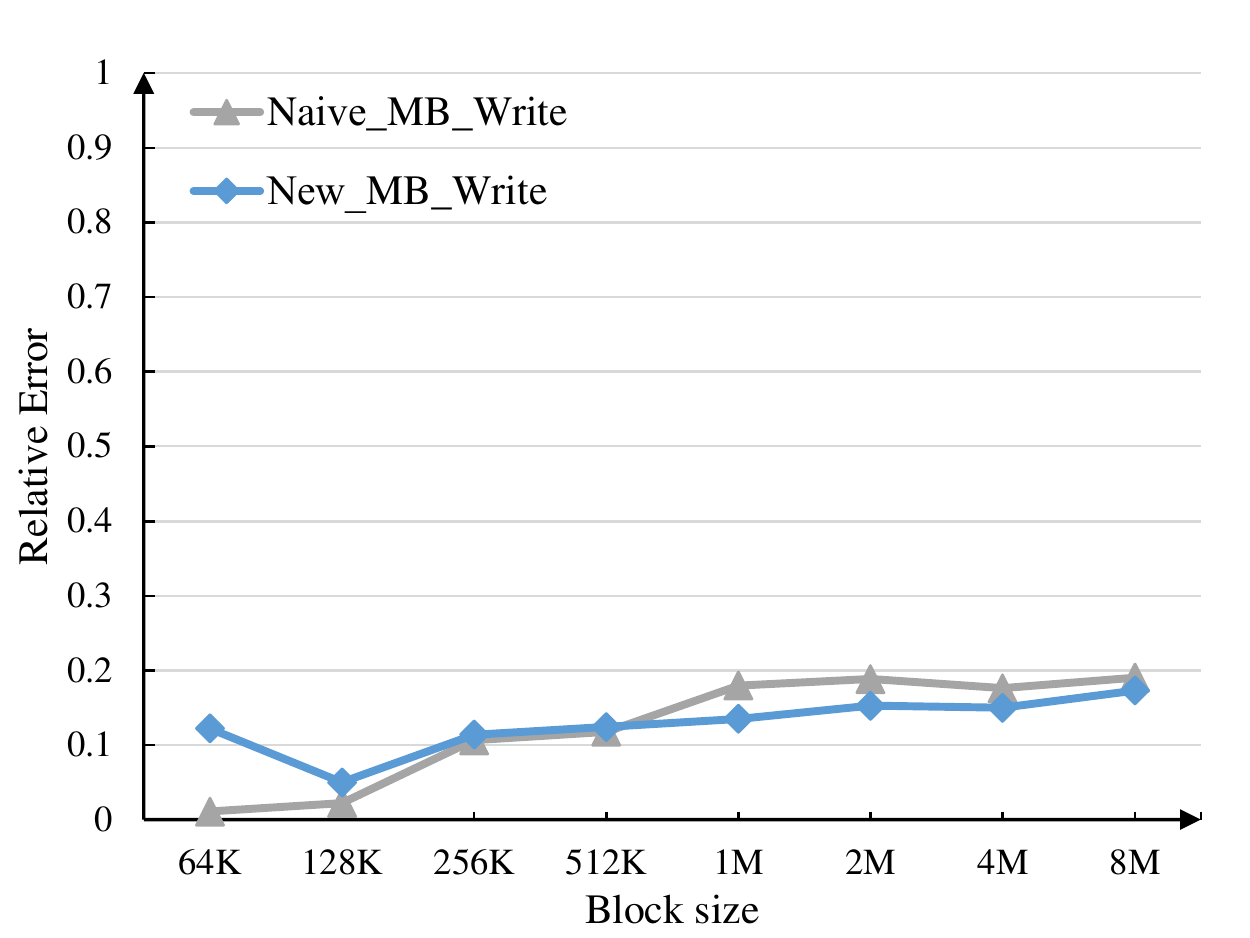}
		\par \centering {\small (a) $t_o$ Relative Error}
		\label{fig:sub1}
	\end{minipage}
	\hfill
	\begin{minipage}[t]{2.3in}
		\centering
		\includegraphics[width=1\linewidth]{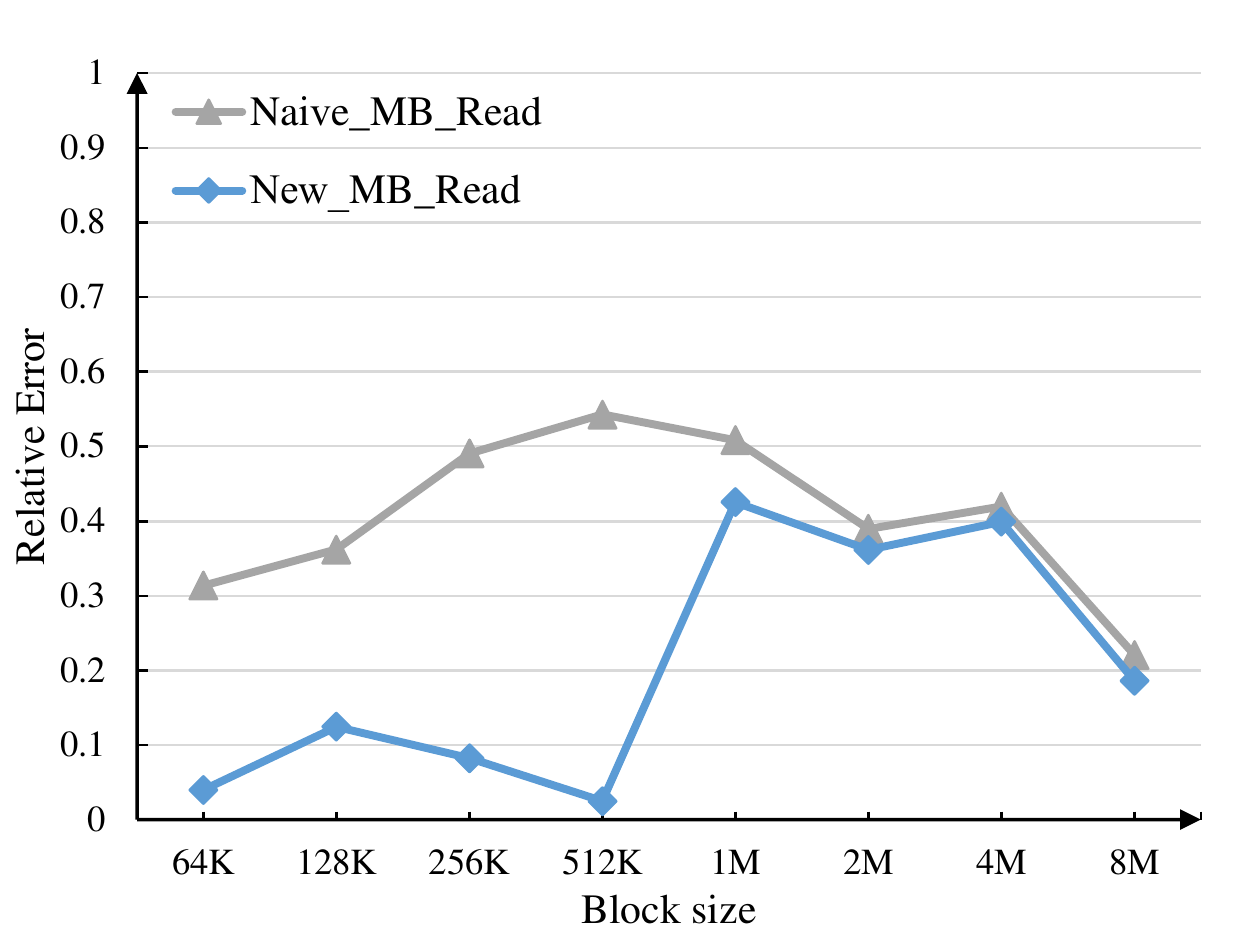}
		\par \centering {\small (b) $t_i$ Relative Error}
		\label{fig:sub2}
	\end{minipage}
	\caption{\small Accuracy of two microbenchmark compared to Real Application with 1 writer and 1 reader on BigRed II.}
	\label{fig:BG2-NaiveMB-MB-Real-1v1}
\end{figure}

The results for two version of microbenchmark with 1 writer and 1 reader are displayed in Figure \ref{fig:BG2-NaiveMB-MB-Real-1v1}.
For the writing relative error, 
the two versions of microbenchmark have similar accuracy among different block sizes. 
%12\% on average
%The naive microbencmark has a relative error of 1\% on 64KB and up to 19\% on 8MB.
%The new microbenchmark has a relative error of 4\% on 128KB and up to 17\% on 8MB.
But for the reading relative error, 
%naive version has a relative error of 22\% on 8MB and up to 54\% on 1MB, 
%and the relative error is 40.5\% on average among different block sizes.
%The new microbenchmark has a relative error of 2\% on 512KB and up to 42\% on 1MB,
%and the relative error is 20.5\% on average among different block sizes.
the new microbenchmark gets better accuracy than naive version on all block sizes.
The relative error of new microbenchmark is 20.5\% on average among different block sizes, while 40.5\% in naive version. 
In this case, the new one has similar accuracy on writing, but it has better accuracy on reading.

\begin{figure}[h]
	\centering
	\begin{minipage}[t]{2.3in}
		\centering
		\includegraphics[width=1\textwidth]{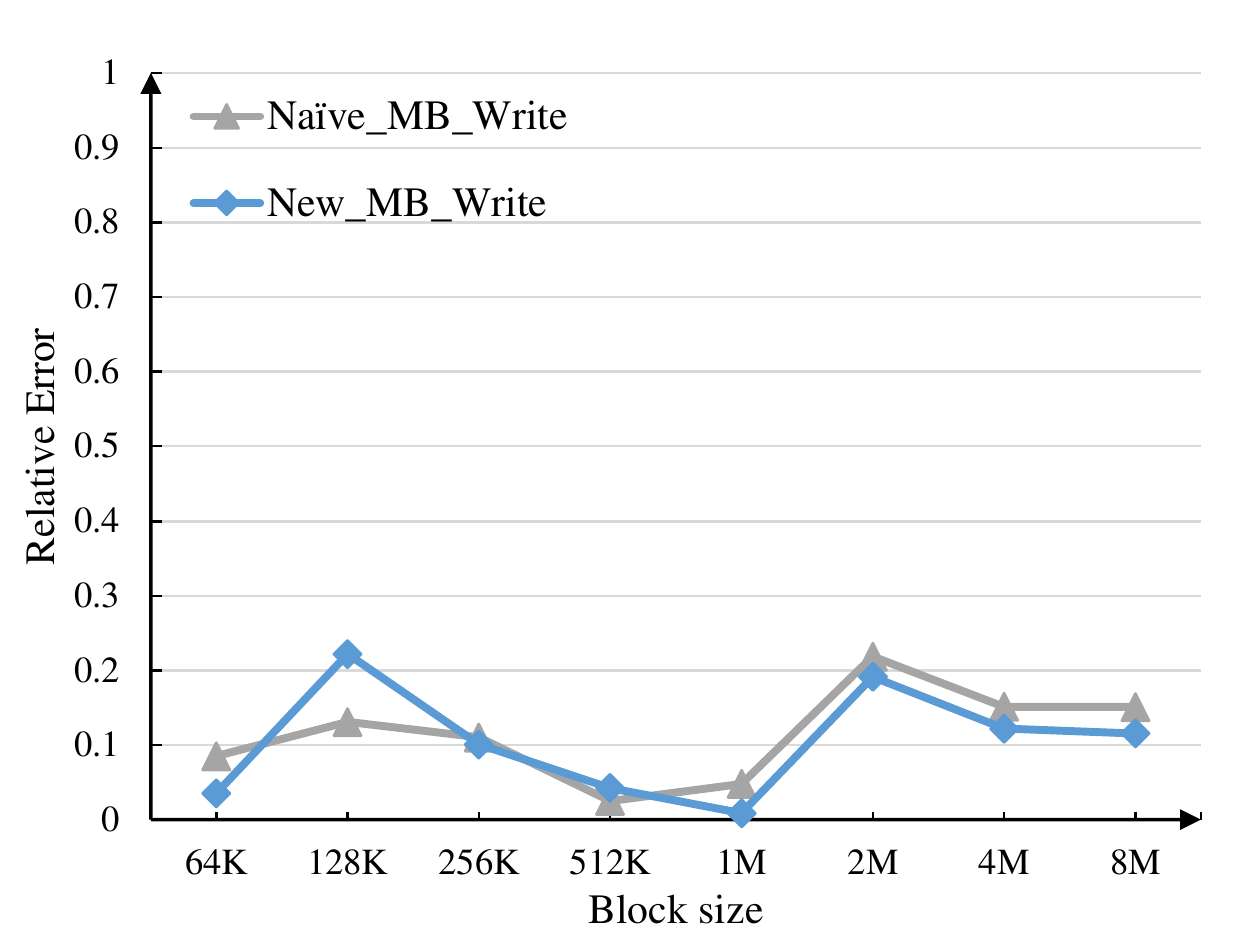}
		\par \centering {\small (a) $t_o$ Relative Error}
		\label{fig:sub1}
	\end{minipage}
	\hfill
	\begin{minipage}[t]{2.3in}
		\centering
		\includegraphics[width=1\linewidth]{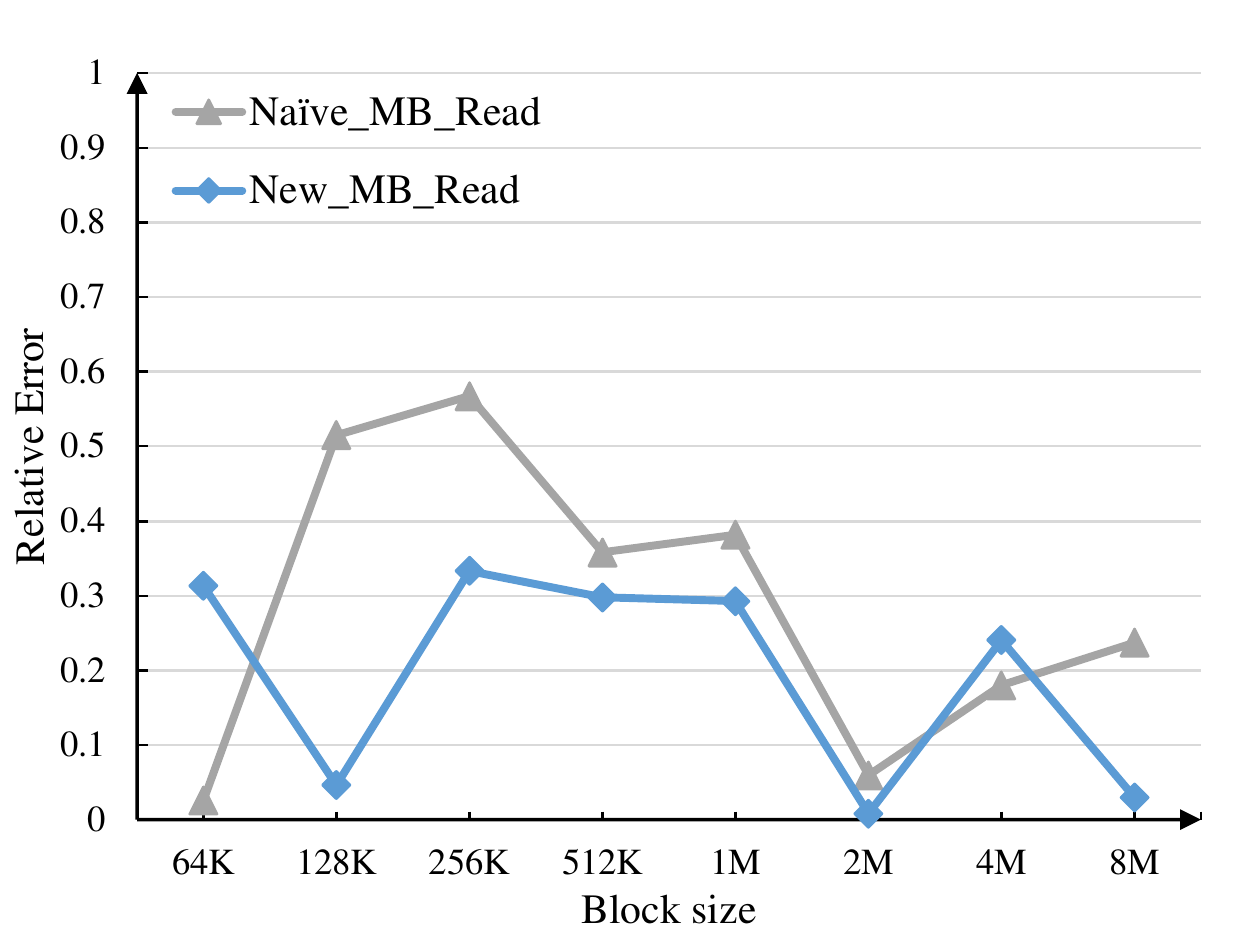}
		\par \centering {\small (b) $t_i$ Relative Error}
		\label{fig:sub2}
	\end{minipage}
	\caption{\small Accuracy of the two microbenchmark compared to Real Application with 4 writers and 1 reader on BigRed II.}
	\label{fig:BG2-NaiveMB-MB-Real-4v1}
\end{figure}

Figure \ref{fig:BG2-NaiveMB-MB-Real-4v1} shows the relative error on writing and reading with 4 writer threads and 1 reader thread.
For the writing relative error, 
the new microbenchmark has better accuracy with 10\% on average among different block sizes, while the naive microbenchmark has 12\% on average.
For the reading relative error, 
the new microbenchmark also obtains better accuracy with 19.5\% on average among different block sizes, while the naive microbenchmark has 29\% on average.
Thus, the new version slightly outperforms naive version on writing, while has higher accuracy on reading.

\begin{figure}[h]
	\centering
	\begin{minipage}[t]{2.3in}
		\centering
		\includegraphics[width=1\textwidth]{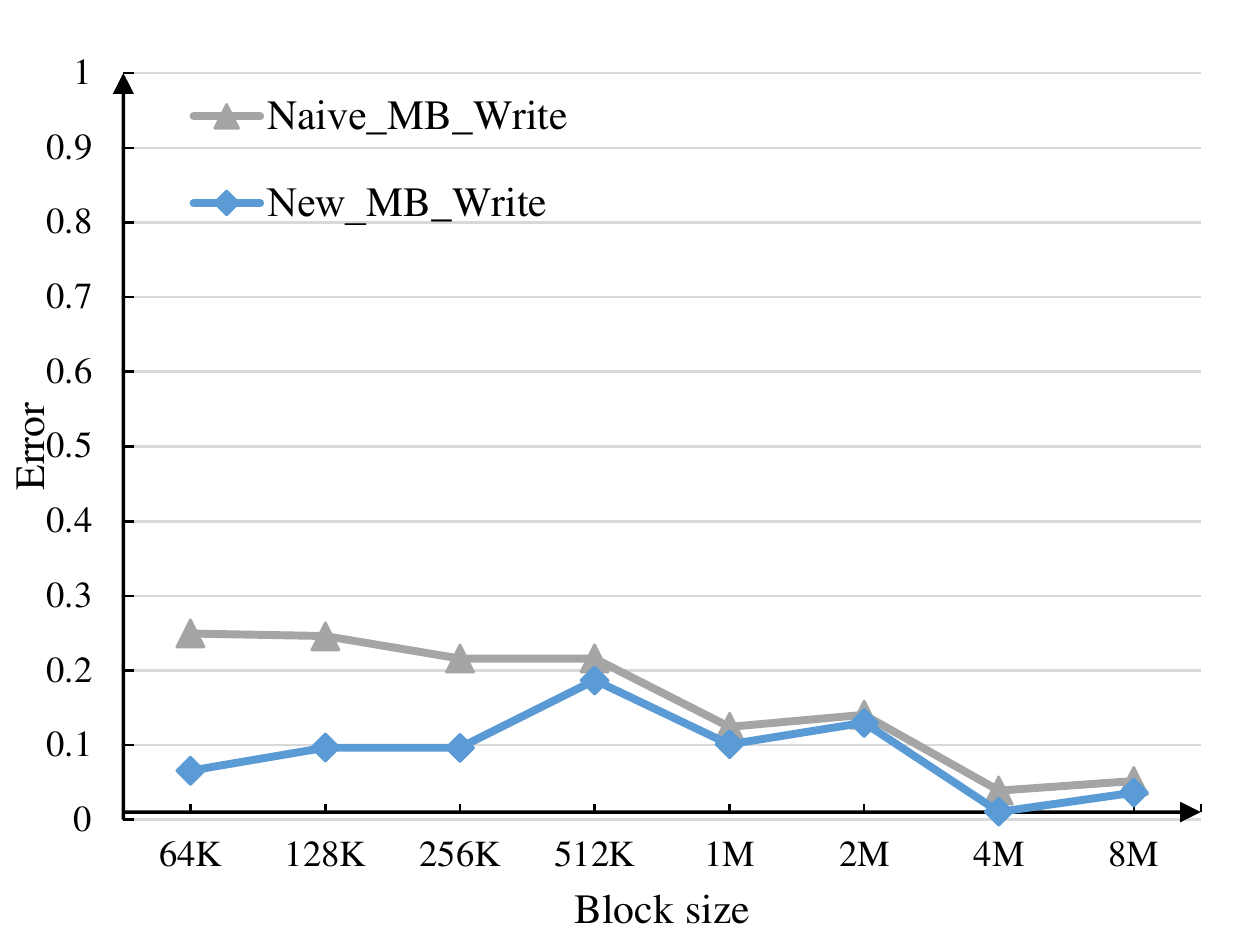}
		\par \centering {\small (a) $t_{o}$ Relative Error}
		\label{fig:sub1}
	\end{minipage}
	\hfill
	\begin{minipage}[t]{2.3in}
		\centering
		\includegraphics[width=1\linewidth]{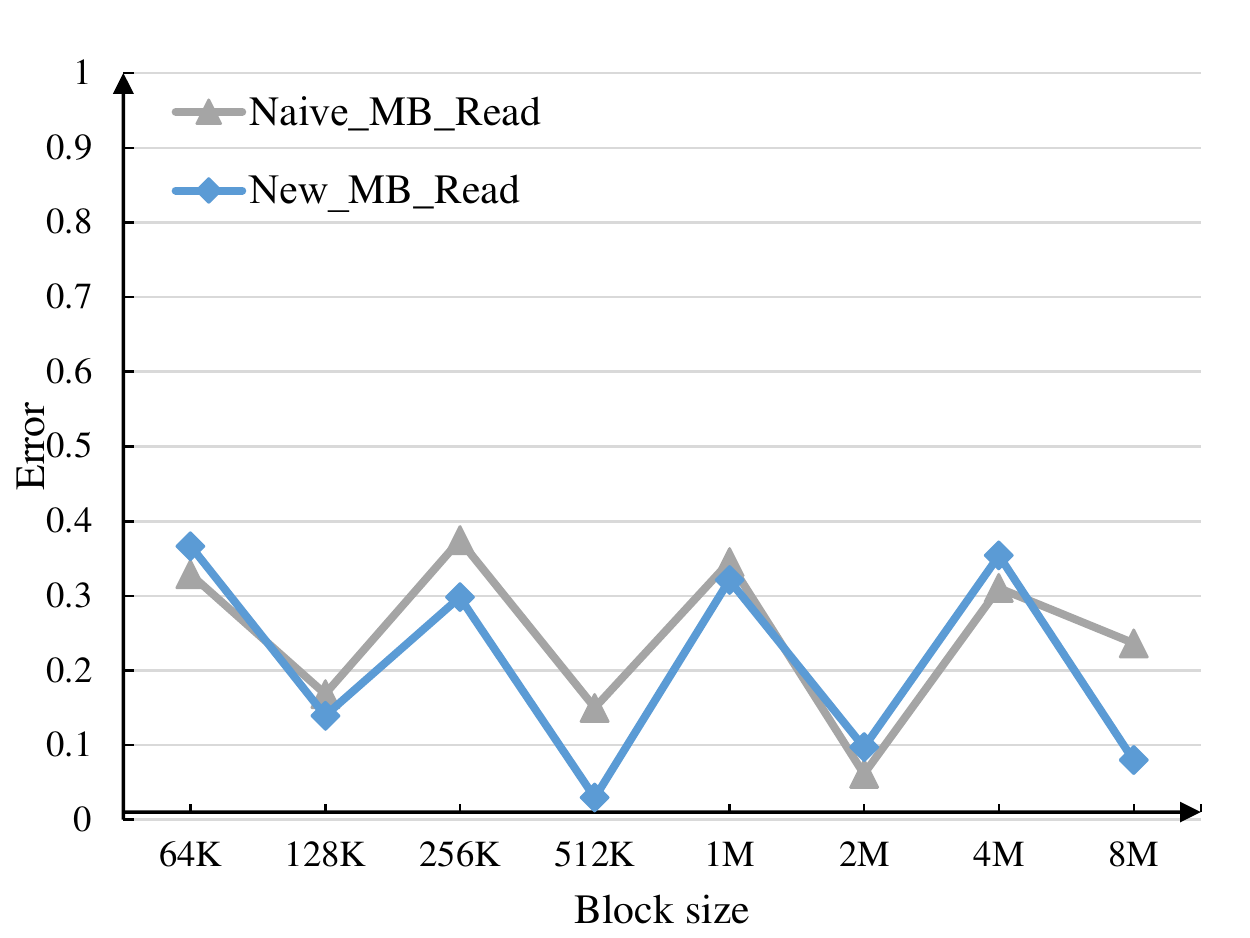}
		\par \centering {\small (b) $t_{i}$ Relative Error}
		\label{fig:sub2}
	\end{minipage}
	\caption{\small Accuracy of two microbenchmark compared to Real Application with 16 writers and 4 readers on BigRed II.}
	\label{fig:BG2-NaiveMB-MB-Real-16v4}
\end{figure}

Figure \ref{fig:BG2-NaiveMB-MB-Real-16v4} shows the relative error on writing and reading with 16 writer threads and 4 reader threads.
For the writing relative error, 
the new microbenchmark has better accuracy with 9\% on average among different block sizes, while the naive microbenchmark has 16\% on average.
%the naive microbencmark has a relative error of 4\% on 4MB and up to 25\% on 64KB, 
%and the relative error is 16\% on average among different block sizes.
%The new microbenchmark has a relative error of 1\% on 4MB and up to 18\% on 512KB,
%and the relative error is 9\% on average among different block sizes.
For the reading relative error, 
the new microbenchmark also obtains better accuracy with 21\% on average among different block sizes, while the naive microbenchmark has 24.6\% on average.
%the naive version has a relative error of 6\% on 4MB and up to 37\% on 256KB,
%and the relative error is 24.6\% on average among different block sizes.
%The new microbenchmark has a relative error of 3\% on 512KB and up to 35\% on 4MB,
%and the relative error is 21\% on average among different block sizes.
In this case, the new version outperforms naive version on both writing and reading accuracy.

\begin{figure}[h]
	\centering
	\begin{minipage}[t]{2.3in}
		\centering
		\includegraphics[width=1\textwidth]{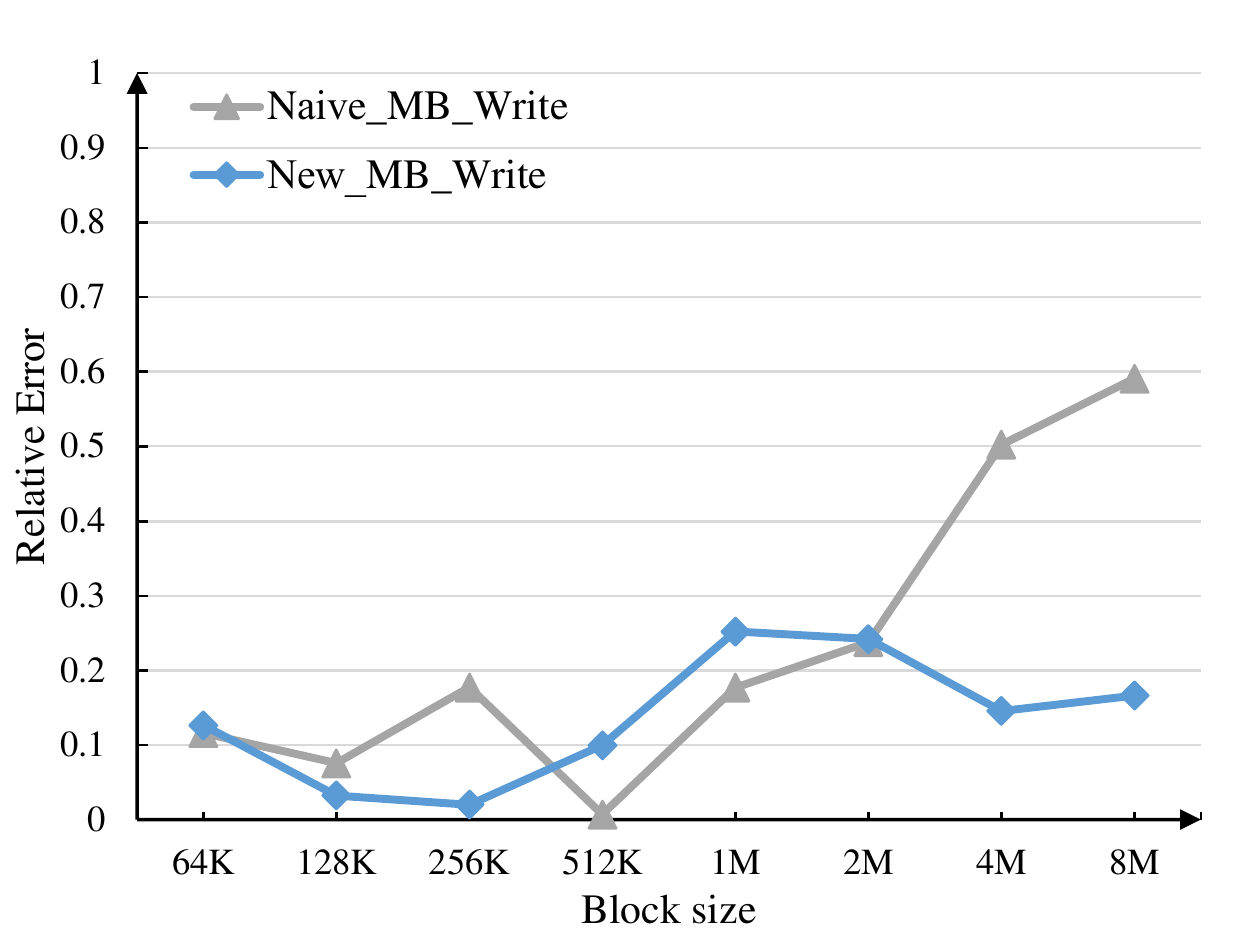}
		\par \centering {\small (a) $t_{o}$ Relative Error}
		\label{fig:sub1}
	\end{minipage}
	\hfill
	\begin{minipage}[t]{2.3in}
		\centering
		\includegraphics[width=1\linewidth]{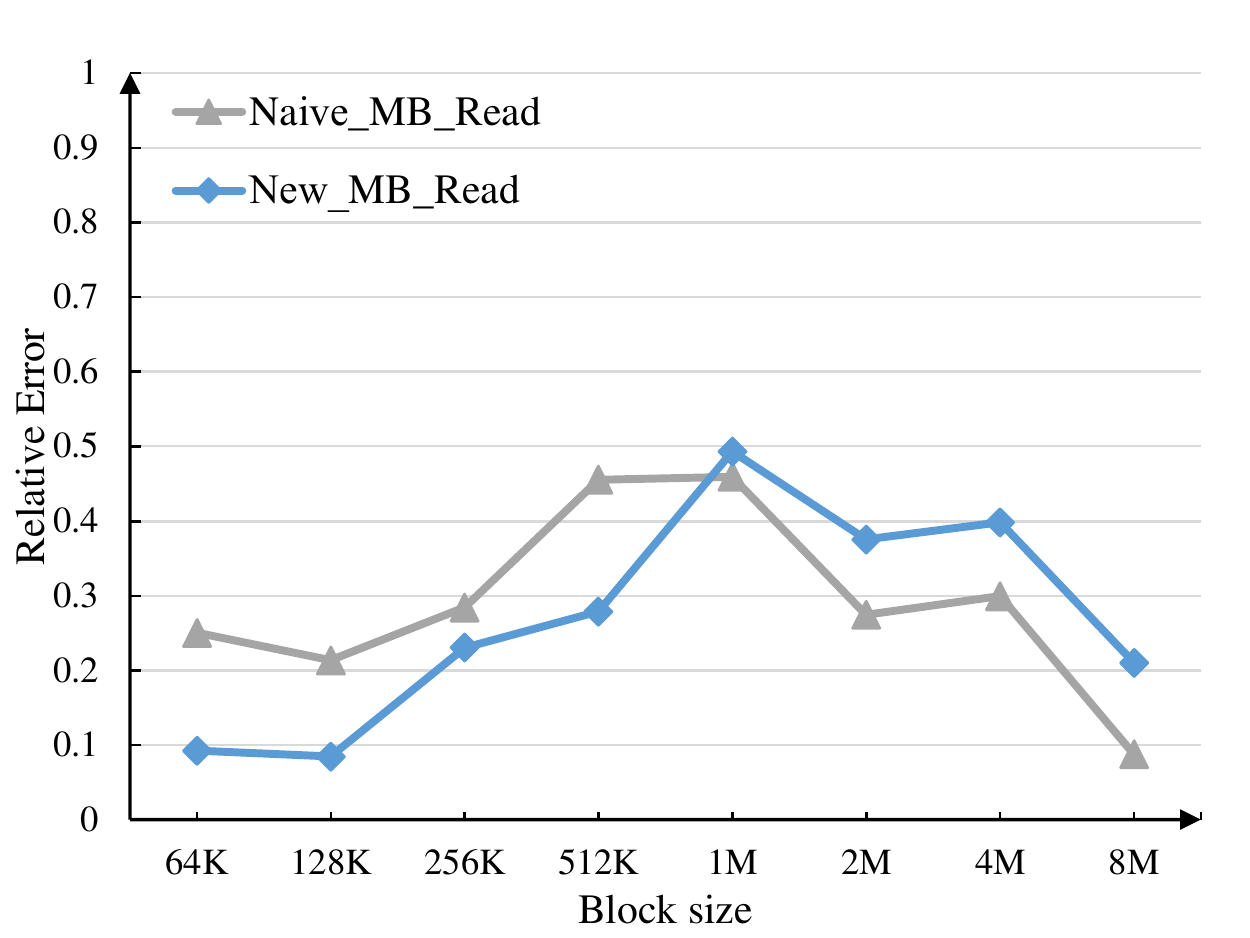}
		\par \centering {\small (b) $t_{i}$ Relative Error}
		\label{fig:sub2}
	\end{minipage}
	\caption{\small Accuracy of two microbenchmark compared to Real Application with 32 writers and 32 readers on BigRed II.}
	\label{fig:BG2-NaiveMB-MB-Real-32v32}
\end{figure}

Figure \ref{fig:BG2-NaiveMB-MB-Real-32v32} shows the relative error on writing and reading with 16 writer threads and 4 reader threads.
For the writing relative error, 
the new microbenchmark has better accuracy with 13.5\% on average among different block sizes, while the naive microbenchmark has 23.4\% on average.
Besides, the naive version even gets a relative error up to 59\% on 8MB.
%the naive microbencmark has a relative error of 1\% on 512KB and up to 59\% on 8MB, 
%and the relative error is 23.4\% on average among different block sizes.
%The new microbenchmark has a relative error of 1\% on 4MB and up to 18\% on 512KB,
%and the relative error is 13.5\% on average among different block sizes.
For the reading relative error, 
the new microbenchmark also obtains better accuracy with 27\% on average among different block sizes, while the naive microbenchmark has 29\% on average.
%the naive version has a relative error from 8.7\% on 8MB to 45.9\% on 1MB,
%and the relative error is 29\% on average among different block sizes.
%The new microbenchmark has a relative error from 8.4\% on 128KB to 49\% on 4MB,
%and the relative error is 27\% on average among different block sizes.
Thus, in this case, the new version also outperforms the naive one on both writing and reading accuracy.

%Discussion:
From the above experiment results, we observe that
the new microbenchmark never gets a relative error of more than 50\%.
Besides, it gets an average relative error rate up to 13.5\% for writing and up to 25\% for reading among all cases and different threads configuration. 
Thus, we can conclude that the new version is more accurate than the naive version.
The reason that the microbenchmark can not sometimes predict the $t_{o}$ and $t_{i}$ accurately is that we do not consider network contention 
and sharing among the parallel file system between users into the microbenchmark model. 
And to add these two parts into the microbenchmark is not a trivial job and we do not expect the microbenchmark too complicated.
But we still can use the microbenchmark to get the trends and acceptable I/O performance to predict $t_{o}$ and $t_{i}$ in real applications.

 %budget: 1.5 pages. total: 6 pgs now

\section{Design and Implementation of DataBroker for the Fully Asynchronous Method}
\label{sect:implementation}
%Our big picture (without detail technique)
%Just block graph framework benefit
%Interface
%Flexibility\\
To enable the fully asynchronous pipeline model,
we design and develop a software prototype called {\em Intelligent DataBroker}.
The interface of the DataBroker prototype is similar to Unix's pipe, which has a writing end and a reading end.
For instance, a computation process will call DataBroker.write(block\_id, void* data) to output data, while
an analysis process will call DataBroker.read(block\_id) to input data.
Although the interface is simple,
it has its own runtime system to provide pipelining, hierarchical buffering, and data prefetching.

\begin{figure}[htbp]
\centering
\includegraphics[width=0.9\textwidth]{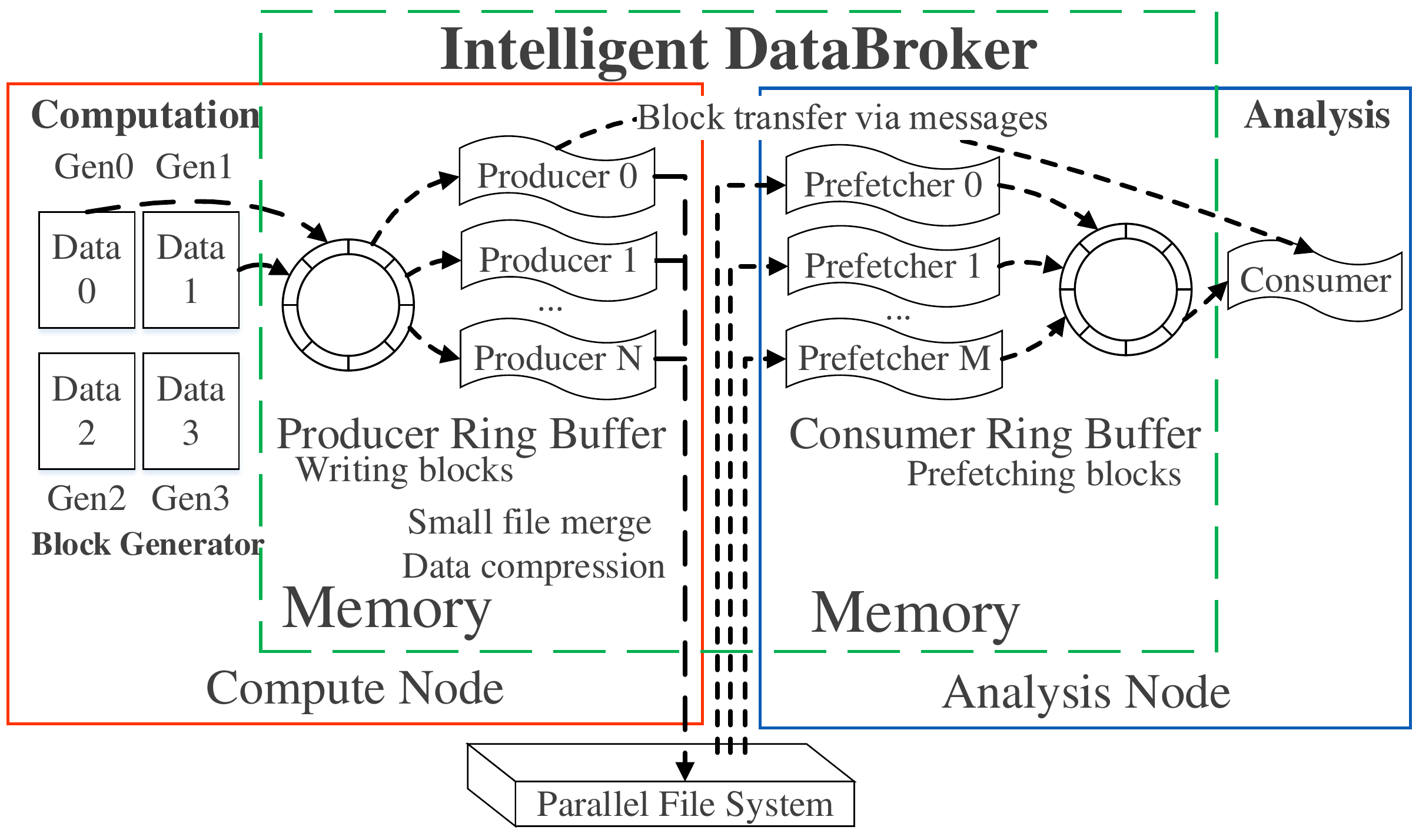}
\caption{\small Architecture of the DataBroker middleware for coupling computation with analysis in a streaming
pipeline manner. DataBroker consists of a producer component on a compute node, and a consumer component on an analysis node.}
\label{fig:big-picture}
\end{figure}

Figure \ref{fig:big-picture} shows the design of DataBroker \cite{fu2016iccs}.
It consists of two components: a {\em DataBroker producer component} in the compute node to
send data, and a {\em DataBroker consumer component} in the analysis node to receive data.
The producer component
owns a producer ring buffer and one or multiple producer threads to process
output in parallel.
Each producer thread looks up the I/O-task queues and uses priority-based scheduling algorithms to
transfer data to destinations in a streaming manner.
%Whenever possible, the data blocks are grouped together and sent out in a batch to maximize the I/O throughput.
A computational process may send data to an analysis process via two possible paths:
message passing by the network, or file I/O by the parallel file system.
%%When and How to use the two paths??
Depending on the execution environment, it is possible that both paths are available and used
to speed up the data transmission time.

The DataBroker consumer is co-located with an analysis process on the analysis node.
The consumer component will receive data from the computation processes, buffer data, and
prefetch and prepare data for the analysis application.
It consists of a consumer ring buffer and one or multiple prefetching threads. %to perform data prefetching.
%Tech1: block-based prefetching
%The prefetching method is particularly designed for fine-grain blocks and the pipeline-style communication
%between computation processes and analysis processes.
The prefetching threads are responsible for making sure there are always
data blocks available in memory by loading blocks from disks to memory.
Since we assume a streaming-based data analysis, the prefetching method can
use the technique of {\it read ahead} to prefetch data efficiently.

 %budget: 1.5 pages.  total: 4.5 pgs now

\section{Experiments with Synthetic and Real Applications}
\label{sect:results}
We perform experiments to verify the accuracy of the analytical model and to evaluate the performance of the fully asynchronous pipeline method, respectively.
For each experiment, we collect performance data from two different programs:
1) a synthetic application, and 2) a real-world computational fluid dynamics (CFD) application.
All the experiments are carried out on BigRed II (a Cray XE6/XK7 system) configured with the Lustre 2.1.6 distributed parallel file system at Indiana University.
Every compute node on BigRed II has two AMD Opteron 16-core Abu Dhabi CPUs and 64 GB of memory,
and is connected to the file system via 56-Gb FDR InfiniBand which is also connected to the DataDirect Network SFA12K storage controllers.

%We conducted two sets of experiments: synthetic experiments and real-world computational fluid dynamics (CFD) application experiments to verify the accuracy of the analytical model and compare the performance among the traditional workflow, the improved version of traditional method, and the fully asynchronous pipeline approach.
%Both kinds of experiments use 32 compute processes communicating with 32 analysis processes. Each process is run on one node.
\subsection{Synthetic and Real-World Applications}
The synthetic application consists of a computation stage and an analysis stage.
To perform these experiments, we use 32 compute nodes to execute the computation stage,
and use two different numbers of analysis nodes (i.e., 2 analysis nodes and 32 analysis nodes) to execute
the analysis stage, respectively.
We launch one process per node.
Each computation process randomly generates a total amount of 1GB data (chopped to small blocks)
and writes the data to the DataBroker producer.
Essentially, the computation processes only generate data, but not perform any computation.
At the same time, each analysis process reads data from its local DataBroker consumer and computes
the sum of the square root of the received data block for a number of iterations.
The mapping between computation processes and analysis processes is static.
For instance, if there are 32 computation processes and 2 analysis processes, each analysis process
will process data from a half of the computation processes.

Our real-world CFD application, provided by the Mathematics Department at IUPUI \cite{zld-lbm},
computes the 3D simulations of flow slid of viscous incompressible fluid flow
at 3D hydrophobic microchannel walls using the lattice Boltzmann method \cite{guo-lbm, Song-icpp15}.
%of a rectangular flexible sheet fastened
%in the middle line and immersed in a fluid flow \cite{zld-immersed, Song-icpp15}.
%The simulation is able to show the complex interactions between an elastic object and the fluid.
This application is written in ANSI C and MPI. %using the immersed boundary method (IB) \cite{Peskin02}
%which is based on the 3-D lattice Boltzmann method (LBM).
We replaced all the file write functions in the CFD application by our DataBroker API.
The CFD simulation is coupled with an data analysis stage, which
computes a series of statistical analysis functions
at each fluid region %(e.g., n-th moments and probability density function)
for every time step of the simulation.
Our experiment takes as input a 3D grid of $512 \times 512 \times 256$, which is distributed to
different computation processes. %So each process will get $128 \times 128 \times 128$ grid.
Similar to the synthetic experiments, we also run 32 computation processes on 32 compute nodes
while running different numbers of analysis processes.
For each experiment, we execute it four times and display their average in our experimental results.

\subsection{Accuracy of the Analytical Model}
We experiment with both the synthetic application and the CFD application to verify the analytical model.
%when a combination of computation and analysis is executed in the fully asynchronous execution manner.
Our experiments measure the end-to-end time-to-solution on different block sizes ranging from 128KB to 8MB.
The experiments are designed to compare the time-to-solution estimated by the analytical model with the actual time-to-solution
to show the model's accuracy.

\begin{figure}[htbp]
\centering
\begin{minipage}[t]{2.3in}
  \centering
  \includegraphics[width=1\textwidth]{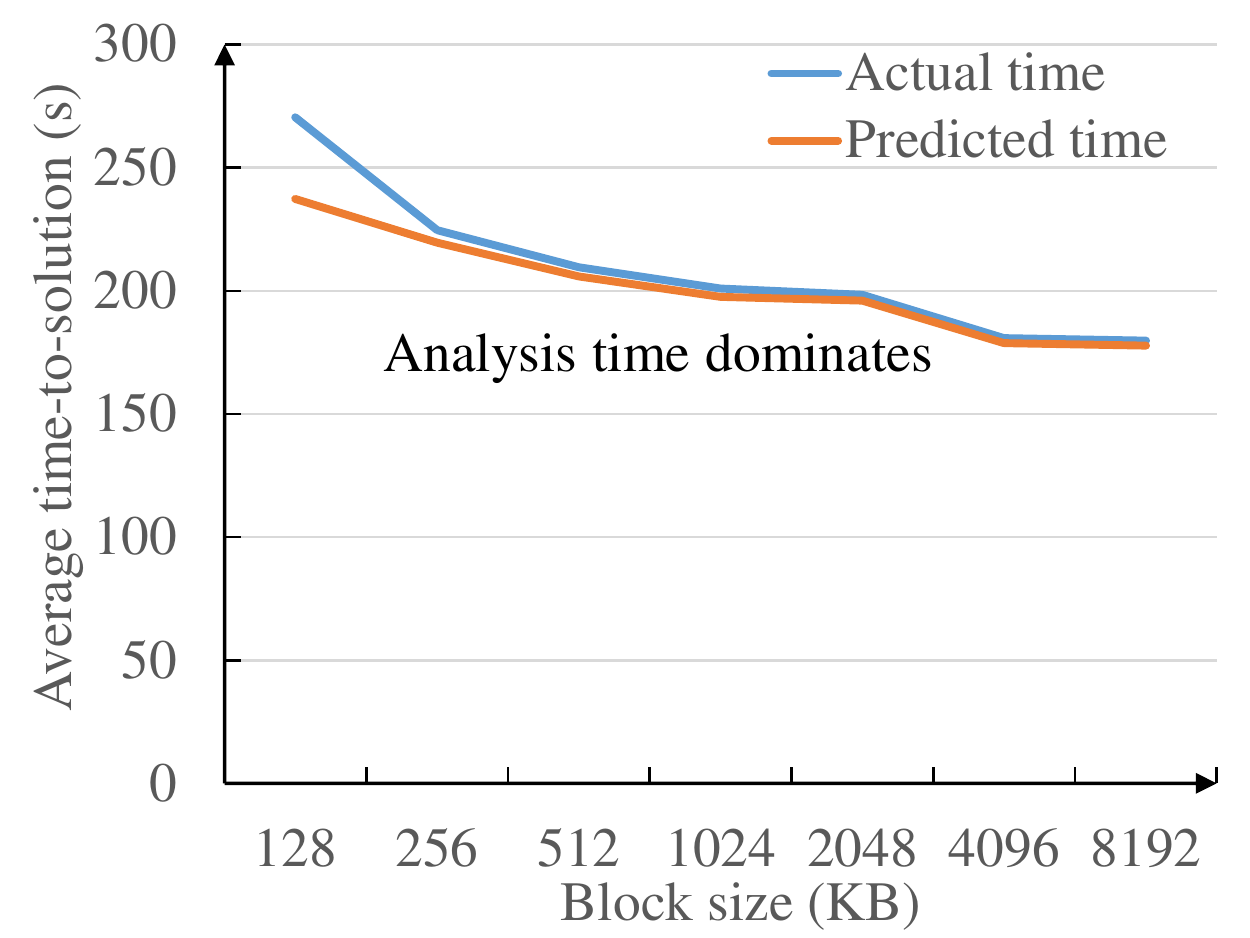}
  \par \centering {\small (a) Synthetic experiments}
  \label{fig:sub1}
\end{minipage}
\hfill
\begin{minipage}[t]{2.3in}
  \centering
  \includegraphics[width=1\linewidth]{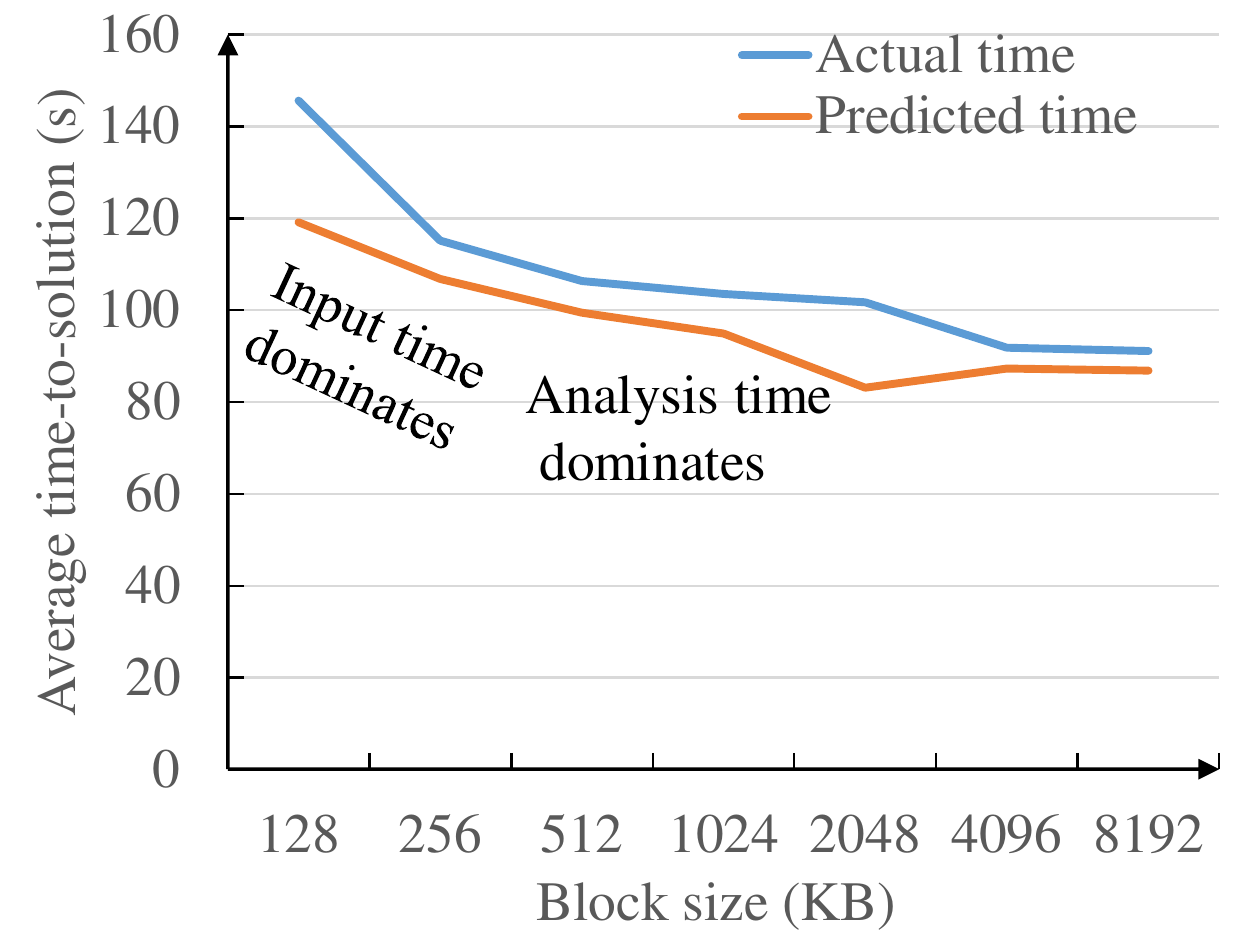}
  \par \centering {\small (b) CFD application}
  \label{fig:sub2}
\end{minipage}
\caption{\small Accuracy of the analytical model for the fully asynchronous pipeline execution with 32 compute nodes and 2 analysis nodes.}
\label{fig:model-verification-32-2}
\end{figure}

Figure \ref{fig:model-verification-32-2} (a) shows the actual time and the predicted time of the synthetic
application using 32 compute nodes and 2 analysis nodes \cite{fu2016iccs}.
For all different block sizes, the analysis stage is the largest bottleneck among the four stages (i.e.,
computation, output, input, and analysis). Hence, the time-to-solution is essentially equal to the analysis time.
Also, the relative error between the predicted and the actual execution time is from 1.1\% to 12.2\%, and on average 3.9\%.
Figure \ref{fig:model-verification-32-2} (b) shows the actual time and the predicted time for the CFD application \cite{fu2016iccs}.
Different from the synthetic application, its time-to-solution is initially dominated by the input time
when the block size is 128KB, then it becomes dominated by the analysis time from 256KB to 8MB.
The relative error of the analytical model is between 4.7\% and 18.1\%, and on average 9.6\%.

The relative error is greater than zero because our analytical model ignores the pipeline startup and drainage time,
and there is also a small amount of pipeline idle time and jitter time during the real execution.
Please note that each analysis process has to process the computed results from 16 computation processes.

\begin{figure}[htbp]
\centering
\begin{minipage}[t]{2.3in}
  \centering
  \includegraphics[width=1\textwidth]{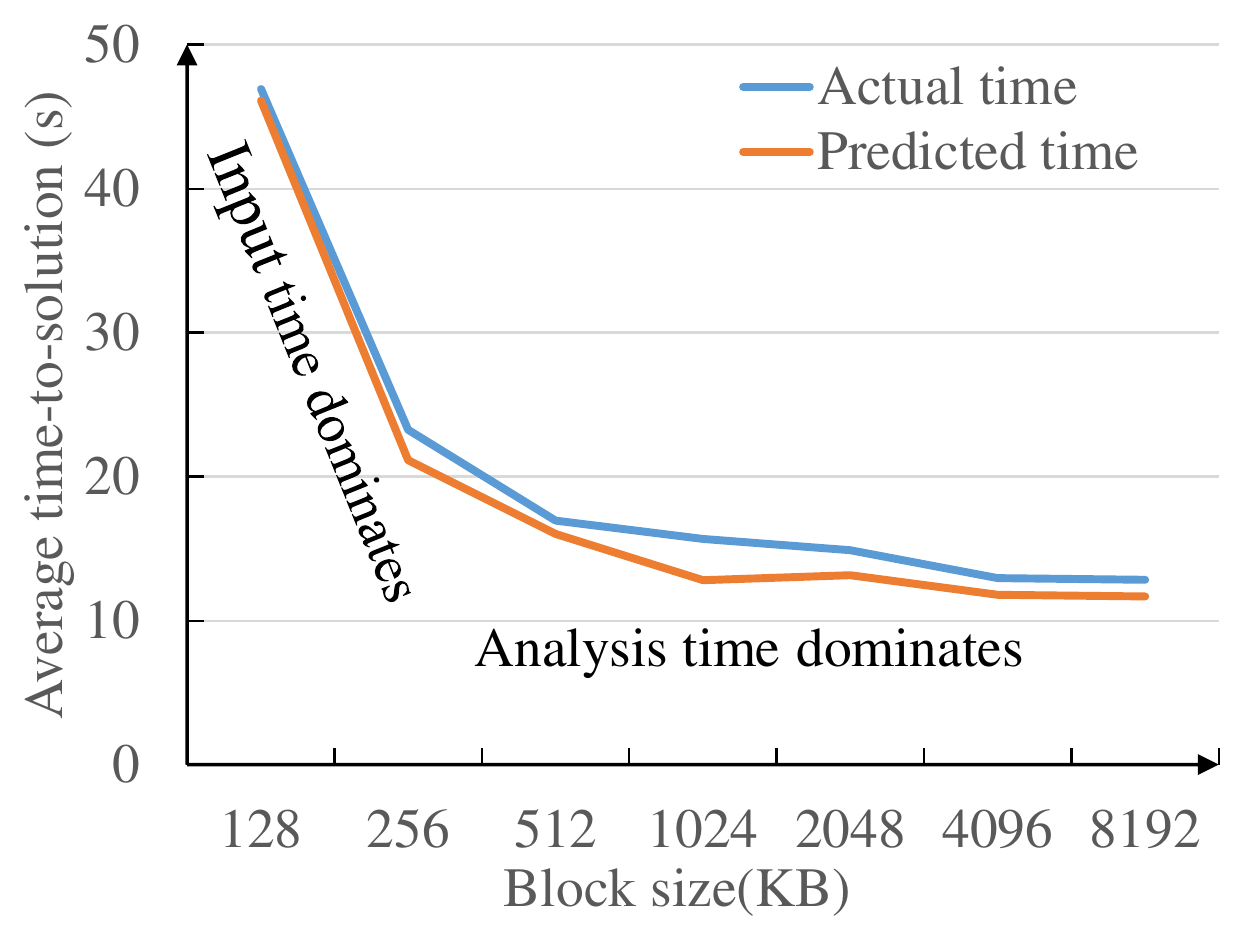}
  \par \centering {\small (a) Synthetic experiments}
  \label{fig:sub1}
\end{minipage}
\hfill
\begin{minipage}[t]{2.3in}
  \centering
  \includegraphics[width=1\linewidth]{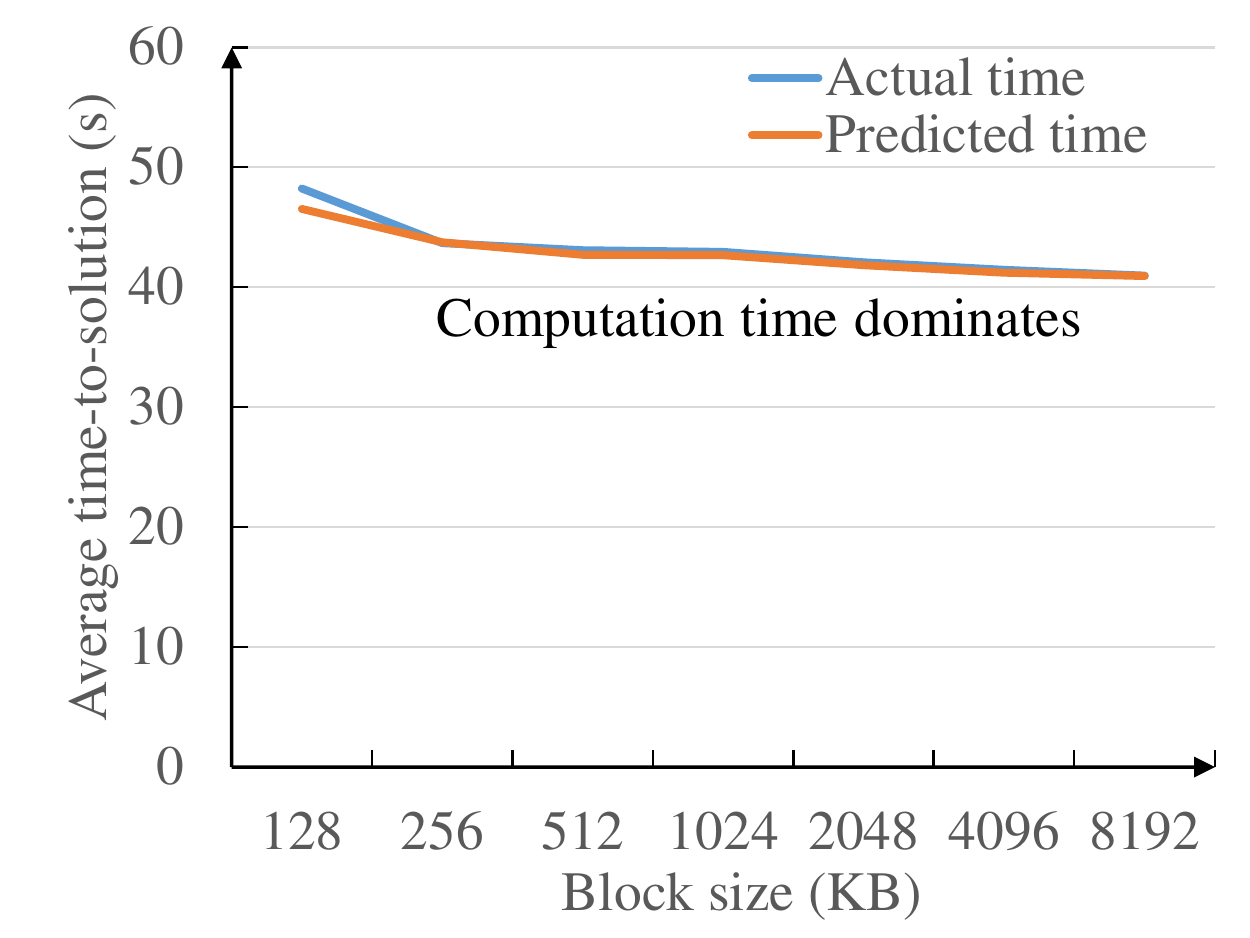}
  \par \centering {\small (b) CFD application}
  \label{fig:sub2}
\end{minipage}
\caption{\small Accuracy of the analytical model for the fully asynchronous pipeline execution with 32 compute nodes and 32 analysis nodes.}
\label{fig:model-verification-32-32}
\end{figure}

%The 32-32 data is new!
Figure \ref{fig:model-verification-32-32} (a) shows the
performance of the synthetic application that uses 32 compute nodes and 32 analysis nodes \cite{fu2016iccs}.
When the block size is equal to 128KB, the input time dominates the time-to-solution.
When the block size is greater than 128KB, the data analysis time starts to dominate the time-to-solution.
%This explains the reason that both curses have a turning point at 256KB.
The turning point in the figure also verifies the bottleneck switch (from the input stage to the analysis stage).
The predicted time and the actual time
are very close to each other and have an average relative error of 9.1\%. %(ranging from 1.75\% to 18.16\%).
Similarly,
Figure \ref{fig:model-verification-32-32} (b) shows an relative error of 0.9\% %(ranging from 0.067\% to 3.56\%).\cite{fu2016iccs}
for the CFD application that also uses 32 compute nodes and 32 analysis nodes \cite{fu2016iccs}.

%Commented out the 32:32 figure!!

\subsection{Performance Speedup}
Besides testing the analytical model,
we also conduct experiments to evaluate the performance improvement by using
the fully asynchronous pipeline method.
The experiments compare three different approaches (i.e., three implementations)
to executing the integrated computation and analysis:
%(as described in Section \ref{sect:model}):
1) the traditional method,
2) the improved version of the traditional method which builds upon fine-grain blocks and
overlaps computation with data output,
and 3) the fully asynchronous pipeline method based on DataBroker.
Each of the three implementations takes the same input size and is compared with each other
in terms of wall clock time.

\begin{figure}[htbp] %  figure placement: here, top, bottom, or page
\centering
\begin{minipage}[t]{2.3in}
  \centering
  \includegraphics[width=1\textwidth]{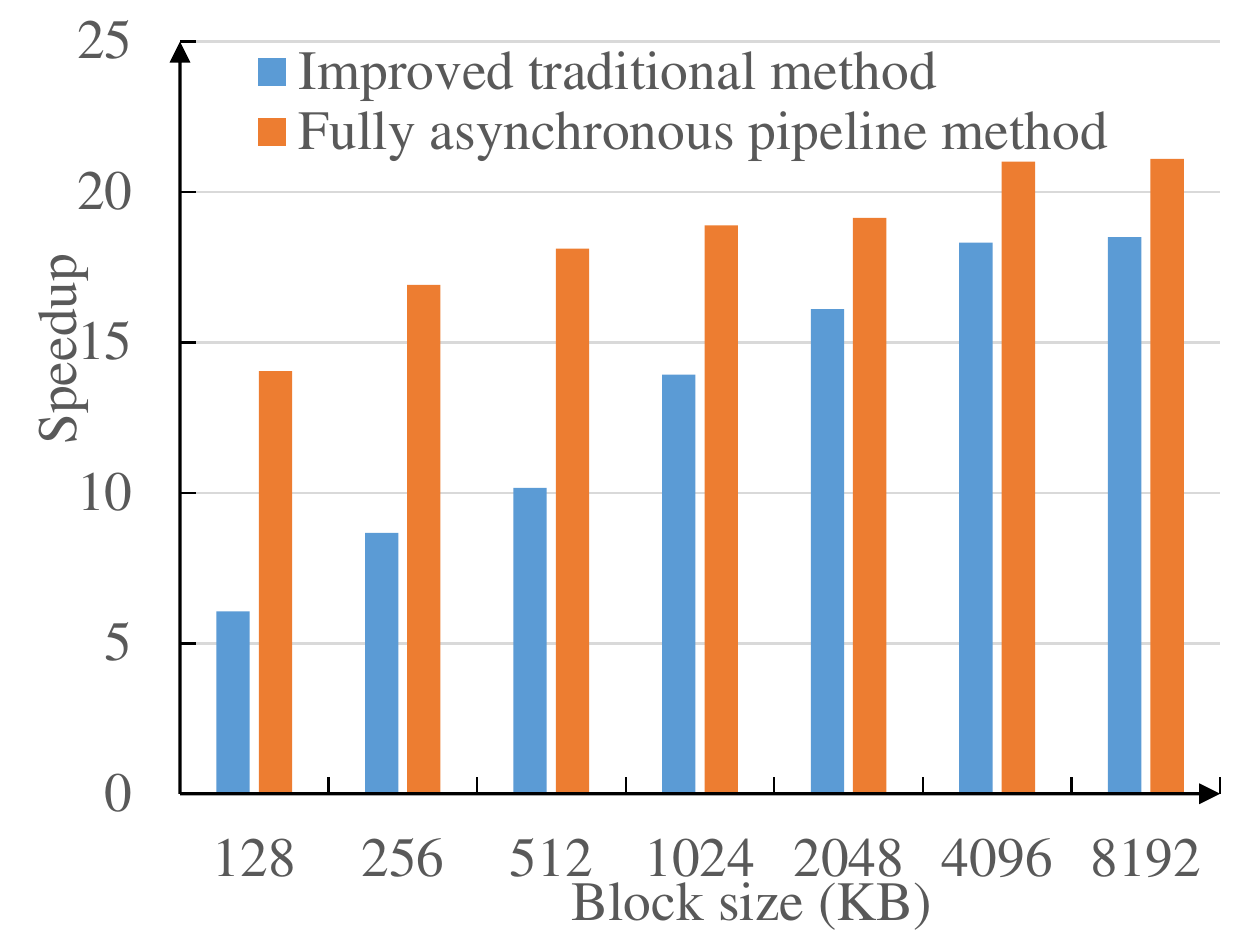}
  \par \centering {\small (a) Synthetic experiments}
  \label{fig:sub1}
\end{minipage}
\hfill
\begin{minipage}[t]{2.3in}
  \centering
  \includegraphics[width=1\textwidth]{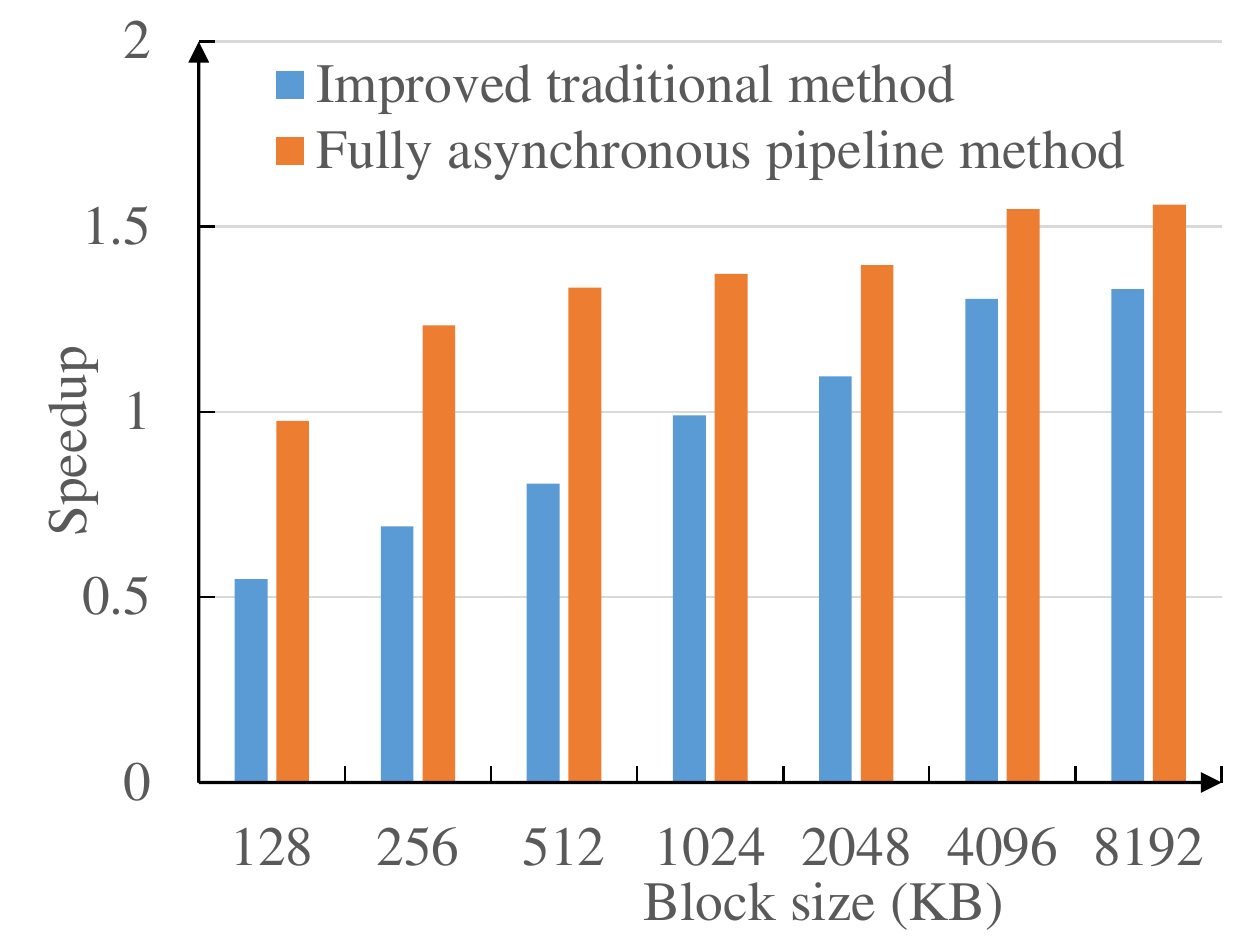}
  \par \centering {\small (b) CFD application}
  \label{fig:sub2}
\end{minipage}
\caption{\small Performance comparison between the traditional, the improved, and the DataBroker-based fully asynchronous methods using 32 compute nodes and 2 analysis nodes.}
%(The traditional method is the baseline program).}
\label{fig:performance-evaluation-32-2}
\end{figure}

Figure \ref{fig:performance-evaluation-32-2} (a) and (b) show the speedup of
the synthetic application and the real-world CFD application, respectively \cite{fu2016iccs}.
Note that the baseline program is the traditional method (i.e., speedup=1).
The data in Figure \ref{fig:performance-evaluation-32-2} (a) shows that the improved version of the traditional method
can be up to 18 times faster than the traditional method when the block size is equal to 8MB \cite{fu2016iccs}.
It seems to be surprising, but
by looking into the collected performance data,
we discover that reading two 16GB files by two MPI process simultaneously
is 59 times slower than reading a collection of small 8MB files by the same two MPI processes.
This might be because two 16GB files are allocated to the same storage device, while a number of 8MB files
are distributed to multiple storage devices.
%reading a total amount of 32GB data (i.e., two MPI processes, each reads a 16GB file) can be 59 times slower than reading a collection of smaller blocks of 8MB asynchronously (also from two MPI processes).
On the other hand, the fully asynchronous pipeline method is faster than the improved traditional method by up to
131\%  when the block size is equal to 128KB.
%We can see that both improved traditional method and the fully asynchronous pipeline begin to have better performance than traditional method with block size 128KB. The fully asynchronous pipeline is 21.1 times faster than traditional method with block size 8MB at most. It can also be 53\% faster than partial pipeline at most with block size 512KB. That is because the block algorithm works. The I/O time for smaller blocks is faster than one whole block.
Figure \ref{fig:performance-evaluation-32-2} (b) shows the speedup of the CFD application \cite{fu2016iccs}.
We can see that the fully asynchronous method is always faster (up to 56\%) than the traditional
method whenever the block size is larger than 128KB. The small block size of 128KB does not lead to
improved performance because writing small files to disks can incur significant file system overhead
and cannot reach the maximum network and I/O bandwidth.
%On block sizes of 4MB and 8MB, the fully asynchronous method can outperform the traditional method by 56\%.
Also, the fully asynchronous method is consistently faster
than the improved traditional method by 17\% to 78\%.

%that both improved traditional method and the fully asynchronous pipeline begin to have better performance than traditional method with block size 256KB. The fully asynchronous pipeline is 56\% times faster than traditional method with block size 8MB at most. It can also be 8 times faster than partial pipeline at most with block size 512KB.
%Figure \ref{fig:performance evaluation} (b) shows the speedup of real world experiment. We can see that fully asynchronous pipeline begin to have better performance than traditional method with block size 128KB. It can be 31\% times faster than traditional method at most with block size 8MB. It can also be 12\% faster than partial pipeline at most with block size 256KB. The reason why it cannot be faster than partial pipeline with bigger block size is that as follows. Our proposed method could save the time in input stage. But in this specific experiment, analysis stage becomes the main bottleneck and much longer than input stage. Thus the total time equals to analysis stage.

\begin{figure}[htbp] %  figure placement: here, top, bottom, or page
\centering
\begin{minipage}[t]{2.3in}
  \centering
  \includegraphics[width=1\textwidth]{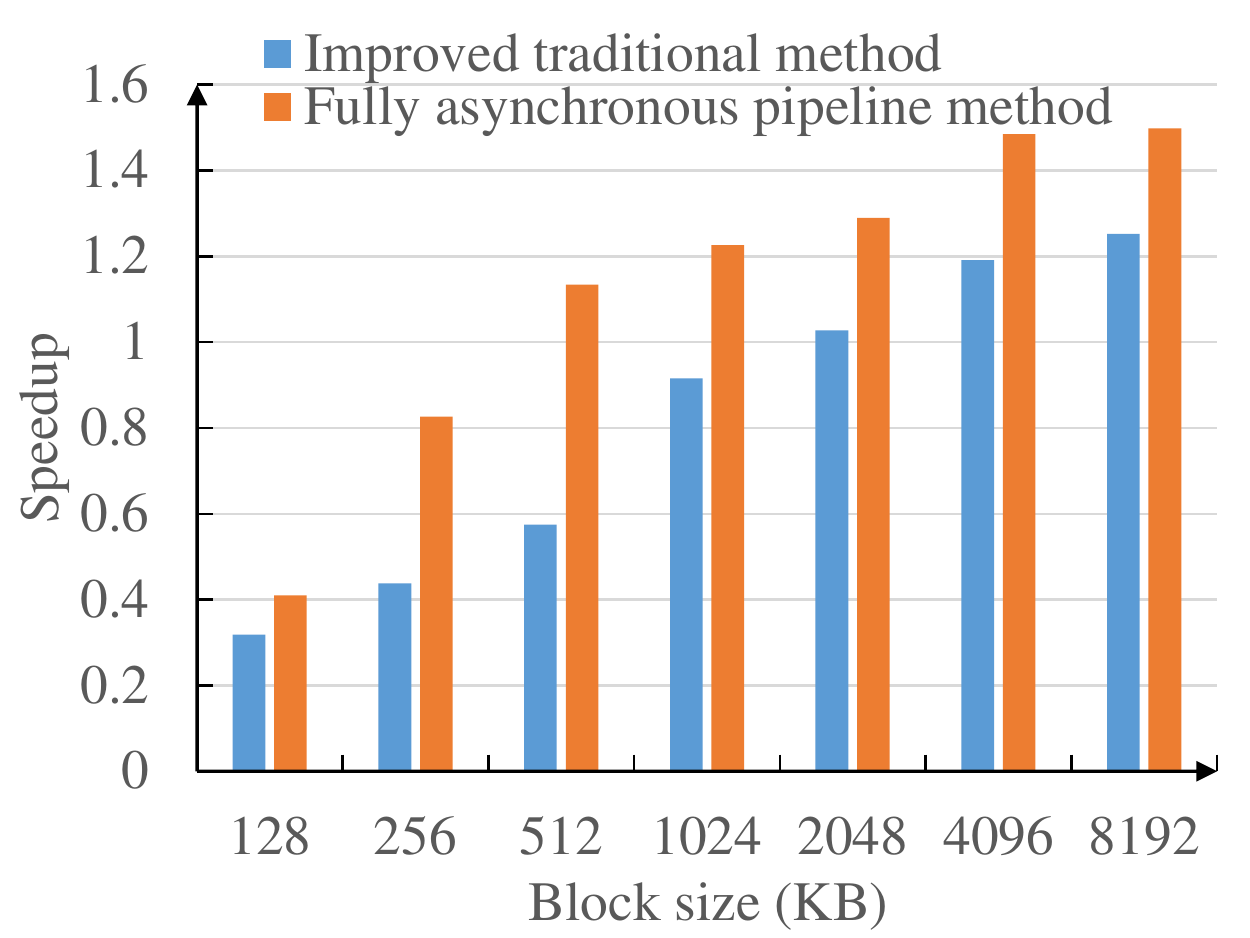}
  \par \centering {\small (a) Synthetic experiments}
  \label{fig:sub1}
\end{minipage}
\hfill
\begin{minipage}[t]{2.3in}
  \centering
  \includegraphics[width=1\textwidth]{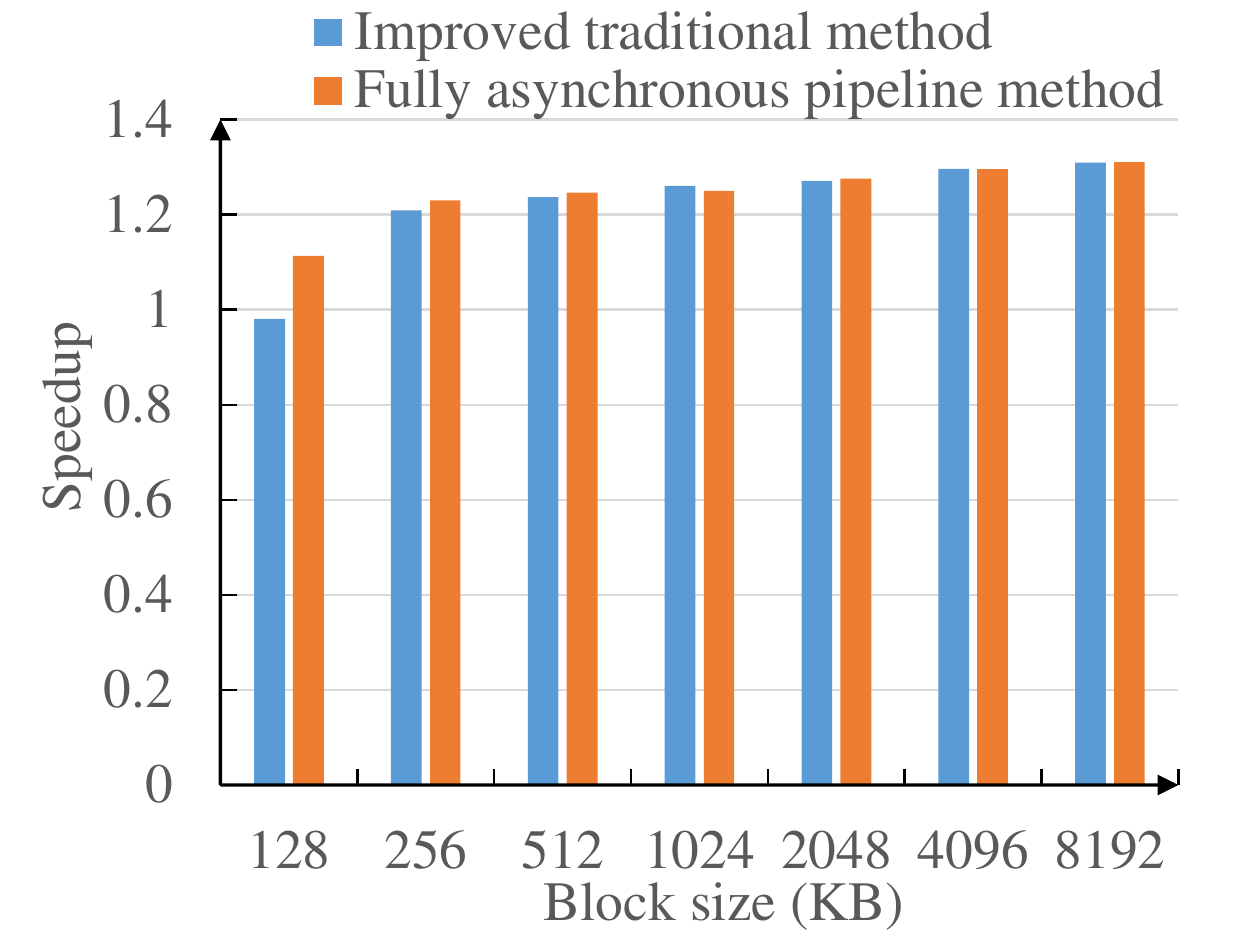}
  \par \centering {\small (b) CFD application}
  \label{fig:sub2}
\end{minipage}
\caption{\small Performance comparison between the traditional, the improved, and the DataBroker-based fully asynchronous methods using 32 compute nodes and 32 analysis nodes.}
\label{fig:performance-evaluation-32-32}
\end{figure}

Figure \ref{fig:performance-evaluation-32-32} (a) shows the speedup of the synthetic
application that uses 32 compute nodes and 32 analysis nodes \cite{fu2016iccs}.
%By comparing traditional method, partial pipeline method, and our proposed method,
We can see that the fully asynchronous pipeline method
is 49\% faster than the traditional method when the block size is equal to 8MB.
It is also 24\% faster than the improved transitional method when the block size is equal to 4MB.
Figure \ref{fig:performance-evaluation-32-32} (b) shows the speedup of the CFD application with 32 compute nodes and 32 analysis nodes \cite{fu2016iccs}.
Both the fully asynchronous pipeline method and the improved traditional method
%when the block size is greater than 128KB.
are faster than the traditional method.
For instance, they are 31\% faster with the block size of 8MB.
%and faster than the improved traditional method by 12\% on the block size of 128KB.
However, the fully asynchronous pipeline method is almost the same as the improved method
when the block size is bigger than 128KB.
This is because the specific experiment's computation time dominates its time-to-solution
so that both methods' time-to-solution is equal to the computation time,
which matches our analytical model.
%This is because the specific experiment's data analysis time is much greater than its data input time
%so that both methods' time-to-solution is equal to the data analysis time.
%The reason why it cannot be faster than partial pipeline with bigger block size is that as follows: fully synchronised method could save the time in the input stage. But in this specific experiment, analysis stage becomes the main bottleneck and much longer than input stage. Thus the total time equals to analysis stage.

 %budget: 2 pages. total: 8 pgs now.

\section{Open Issues and Challenges}
\label{sect:challenges}
%2 pages
This chapter presents a new way to accelerate scientific workflows consisting of computation-intensive applications and data-intensive applications.
Although the current method has shown promising results, there are more open issues and challenging problems. 
Here we provide a list of challenging problems as follows:
\begin{itemize}
\item Is it possible to design an adaptive data transfer method that can utilize both message passing and parallel file systems concurrently at runtime? Considering new file systems will utilize more Nonvolatile Memories (NVM), the performance 
difference between memory and file system becomes lesser.
\item How to utilize the proposed analytical model to schedule computing resources more efficiently? It can lead to more efficient scheduling methods.
\item Is it possible to extend the task-based pipeline approach to a general workflow model that consists of applications in a task graph? 
\item How to build a general-purpose workflow framework and programming tool to automatically combine different applications seamlessly at the fine-grain task level?
\end{itemize}
These open issues and challenges will require further research in SDN to facilitate faster
data transfer and minimized time-to-solution in tightly coupled workflow applications. 
%involving both computation-intensive and data-intensive applications. %budget: 2 pages. total: 8 pgs now.

\section{Conclusion}
\label{sect:conclusion}
To facilitate the convergence of computational modeling/simulation and the big data analysis, in this chapter
we study the problem of integrating computation with analysis in both theoretical and practical ways.
%This work targets optimizing the end-to-end time-to-solution. of scientific discovery.
First, we use the metric of the time-to-solution of scientific discovery
to formulate the integration problem and propose a fully asynchronous pipeline method to model the execution.
Next, we build an analytical model to estimate the overall time to execute the asynchronous combination of computation and analysis.
In addition to the theoretical foundation, we also design and develop
an intelligent DataBroker to help fully interleave the computation stage and the analysis stage.
%The DataBroker provides pipelining, hierarchical buffering, and prefetching to minimize the time-to-solution.

The experimental results show that the analytical model can estimate the time-to-solution with an average relative error of less than 10\%.
By applying the fully asynchronous pipeline model to both synthetic and real-world CFD applications, %using 32 compute nodes and two analysis nodes,
we can increase the performance of the improved traditional method by up to 131\%
for the synthetic application, and up to 78\% for the CFD application.
%Our future work along this line is to extend DataBroker to send blocks from computation processes
%to analysis processes through the faster path of MPI messages and parallel file systems.
%The other future work %is to merge numerous small file writes into one or more large file writes to reduce the I/O overhead and increase the effective I/O bandwidth.
%%The other future work
%is to utilize the analytical model to appropriately
%allocate resources (e.g., CPUs) to the computation, I/O, and analysis stages
%to optimize the end-to-end time-to-solution by eliminating the most dominant bottleneck.

\section*{Acknowledgements}
This material is based upon research partially supported by the Purdue Research Foundation and by the NSF Grant No. 1522554. Development and experiment of the software framework have used the NSF Extreme Science and Engineering Discovery Environment (XSEDE), which is supported by National Science Foundation grant number ACI-1053575. 

\bibliographystyle{abbrv}
\bibliography{paper}

\end{document}